\documentclass[12pt]{iopart}
\usepackage{iopams,amsthm} 
\usepackage{graphicx}

\begin{document}
 
\title{Instability of enclosed horizons}

\author{Bernard S. Kay}

\address{Department of Mathematics, University of York, York YO10 5DD, UK}
 
\ead{({\rm address for correspondence}) bernard.kay@york.ac.uk}

\begin{abstract}
We point out that there are solutions to the scalar wave equation on 1+1 dimensional Minkowski space with finite energy tails which, if they reflect off a uniformly accelerated mirror due to (say) Dirichlet boundary conditions on it, develop an infinite stress-energy tensor on the mirror's Rindler horizon.  We also show that, in the presence of an image mirror in the opposite Rindler wedge, suitable compactly supported arbitrarily small initial data on a suitable initial surface will develop an arbitrarily large stress-energy scalar near where the two horizons cross.   Also, while there is a regular Hartle-Hawking-Israel-like state for the quantum theory between these two mirrors, there are coherent states built on it for which there are similar singularities in the expectation value of the renormalized stress-energy tensor.   We conjecture that in other situations with analogous \textit{enclosed horizons} such as a (maximally extended) Schwarzschild black hole in equilibrium in a (stationary spherical) box or the (maximally extended) Schwarzschild-AdS spacetime, there will be similar stress-energy singularities and almost-singularities -- leading to instability of the horizons when gravity is switched on and matter and gravity perturbations are allowed for.  All this suggests it is incorrect to picture a black hole in equilibrium in a box or a Schwarzschild-AdS black hole as extending beyond the past and future horizons of a single Schwarzschild (/Schwarzschild-AdS) wedge.  It would thus provide new evidence for 't Hooft's \textit{brick wall} model while seeming to invalidate the picture in Maldacena's `\textit{Eternal black holes in AdS}'. It would thereby also support the validity of the author's \textit{matter-gravity entanglement hypothesis} and of the paper `\textit{Brick walls and AdS/CFT}' by the author and Ort\'iz. 
\end{abstract}

\section{Introduction}

One of the difficult aspects of the problem of quantizing gravity is that the spacetime metric (whether we expect it to be a fundamental dynamical variable in its own right or an emergent quantity) will participate in the dynamics.  So the arena in which our quantum dynamics takes place cannot itself be a (fixed) spacetime.  Nevertheless, we often resort, in practice, to picturing some particular classical spacetime or other as our arena.  For example, we talk about some quantum black hole state by referring to, say, the Schwarzschild solution and assume that we can think of the quantum theory in terms of quantum fluctuations about that background.   Similarly, in AdS/CFT \cite{Maldacena1} we often talk, loosely, about the bulk being AdS or, in an example which is a sort of combination of the previous two examples, Schwarzschild-AdS.

In this paper, we wish to address the question:  How should we picture the spacetime of a (stationary, spherically symmetric)  black hole which is \textit{enclosed}? Or, to put the question precisely:  

\smallskip

\noindent
\textit{To the extent that we can describe the quantum state of such a system with a classical spacetime, what should we take that classical spacetime to be?} 

\smallskip

To explain what we mean by `enclosed' here, suffice it to say for the moment (we will elaborate on this later) that we deem an ordinary Schwarzschild black hole to be enclosed if it is placed in a (stationary, spherical) box, say of area $4\pi R^2$ and suitable boundary conditions are put on the wall of that box.  On the other hand, we deem Schwarzschild-AdS to already be enclosed, by its conformal boundary, since localized perturbations reach that boundary in a finite amount of time.  Again, we assume suitable boundary conditions are put on the conformal boundary.  

\begin{figure}
   \centering
    \includegraphics*[scale=0.5,bb=1.5in 3in 9in 9in]{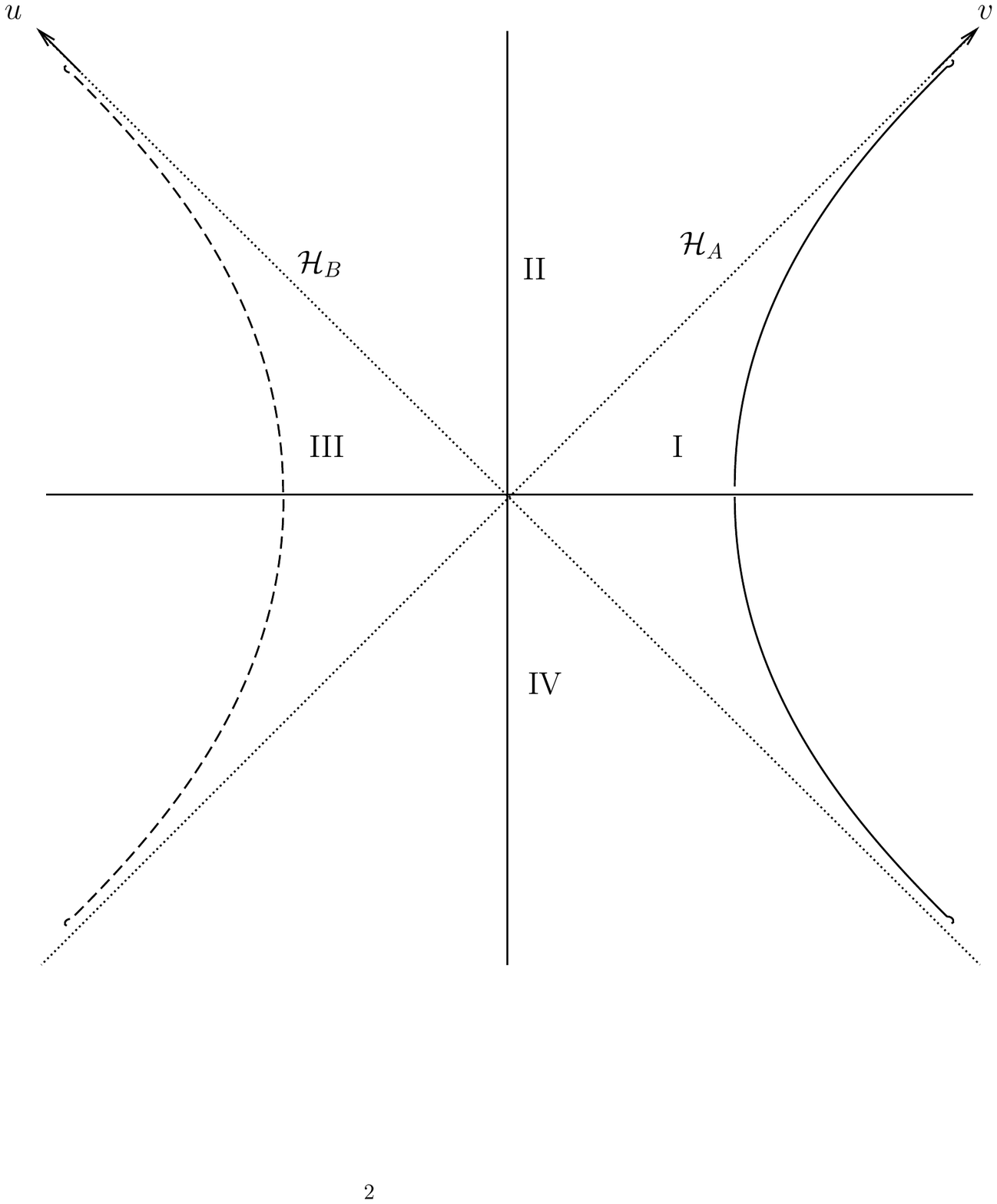}
   \caption{One, seemingly possible, picture for the spacetime of an enclosed stationary spherical black hole (or of the region to the left of an accelerated mirror in Minkowski space).  The right-hand hyperbola represents the box wall or the mirror.  A possible image box wall/image mirror is indicated by the left-hand dashed hyperbola.  ${\cal H}_B$ and ${\cal H}_A$ are the two crossed (full) horizons. When interpreted as Schwarzschild or Schwarzschild-AdS, each point represents a 2-sphere of area
$4\pi r^2$. (In the latter cases, the spacetime is, of course, also bounded by the usual $r=0$ singularities in Regions II and IV which we have omitted from the drawing.) \label{fig1} } 
 \end{figure}

\begin{figure}
   \centering
     \includegraphics*[scale=0.5,bb=2in 3in 7in 9in]{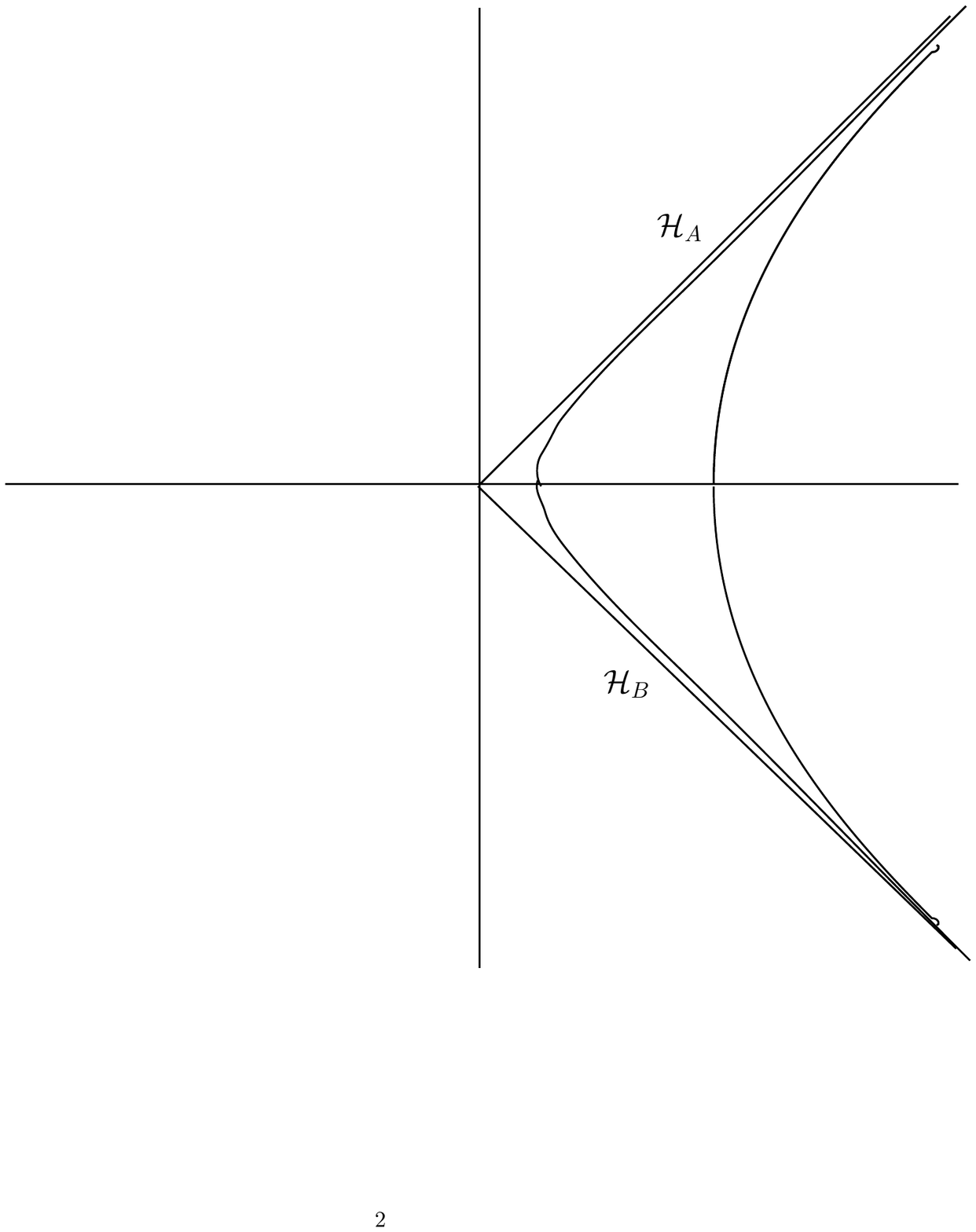}
   \caption{Another possible picture for the spacetime of an enclosed stationary spherical black hole or (see text) of the region to the left of an accelerated mirror in say 1+1 dimensions.   When interpreted as (exterior) Schwarzschild or Schwarzschild-AdS, each point represents a 2-sphere of area $4\pi r^2$. The outer hyperbola represents the location of the box wall/mirror while the region to the left of the inner hyperbola, which includes the future and past (right, half-)horizons, ${\cal H}_A$ and ${\cal H}_B$, is to be interpreted with caution since, as we shall argue, we expect the notion of a 
classical spacetime to break down in this region due to quantum gravity effects.
\label{fig2} }
\end{figure}

Should we picture such an enclosed black hole as in Figure \ref{fig1}, which, in our Schwarzschild-black-hole-in-a-box example, represents the region of the maximal analytic extension to the left of the hyperbola $r=R$ $<$\ref{Note:moot}$>$\footnote{The small Roman numerals in angle-brackets refer to the end section, Section \ref{Sect:notes}, entitled `Notes'.}?   In the case of Schwarzschild-AdS, the undashed and dashed hyperbolae represent, together, the (disconnected) conformal boundary of the maximal analytic extension.  (See e.g.\ \cite{Maldacena2}).

Or should we picture it as in Figure \ref{fig2}, which may be interpreted as the region of the exterior Schwarzschild solution of mass $M$ with $r$-coordinate less than $R$ or the, say, right Schwarzschild wedge of Schwarzschild AdS with mass $M$ -- with a caution that for suitable $\Delta r$ $<$\ref{Note:Deltar}$>$ , the notion of a classical spacetime is expected to break down for $r < 2M+\Delta r$ due to quantum gravity effects?

The first sort of picture is, i.a., assumed in Maldacena's paper \cite{Maldacena2} on the interpretation of the AdS/CFT connection \cite{Maldacena1} in the context of a Schwarzschild-AdS bulk. The second sort of picture is implicit e.g.\ in 't Hooft's \cite{tHooft} \textit{brick-wall model} for a (stationary, spherical) quantum black hole in a box.  (See also \cite{Mukohyama-Israel}.)  

In the present paper, we shall argue for the validity of the second sort of picture and against the first.  In fact we shall argue for some conjectures which entail that, quite generally, enclosed (stationary) horizons, as schematically illustrated in Figure \ref{fig1}, are (both classically and quantum mechanically) unstable.  (As we discuss further at the start of Section \ref{Sect:conjecture}, these conjectures assume that the enclosure-walls/mirrors are asymptotically null as depicted in Figure \ref{fig1} and that the suppressed dimensions in Figure \ref{fig1} are compact.)  These conjectures apply, in particular, to enclosed static, spherically symmetric, black-hole horizons and the instability, if it holds, is therefore distinct from the long-ago-discussed \cite{Press-Teuk} instability that holds for a Kerr black hole in a box due to superradiance.  It is also distinct from the instability of white-hole horizons first proposed and argued for by Eardley \cite{Eardley} in 1974 (see also \cite{BlauGuth, Blau, Lake, WaldRamaswamy}); those horizons belong to white holes which are the time-reverse of black holes formed from stellar collapse, whereas our conjectured instability arises from the presence of white-hole horizons in enclosed (full, time-symmetric) Kruskal and related spacetimes.  

Our argument for these conjectures is mainly based on what happens in the simplest possible example of an enclosed horizon:  namely where we interpret Figure \ref{fig1} as the region of 1+1 dimensional Minkowski space to the left of a uniformly accelerating mirror -- and, we shall usually assume, to the right of another, uniformly decelerating, image mirror represented by the dashed line.  We remark straight away that eternally accelerated mirrors might, of course, be regarded as physically unrealistic.   But our main purpose in studying this system is not to draw any physical conclusions about actual accelerated mirrors (although we believe it likely that such conclusions can be drawn -- see Endnote $<$\ref{Note:realmirrors}$>$).  Rather it is to have a useful analogue system to the Schwarzschild black hole in a box and Schwarzschild-AdS spacetimes depicted in Figure \ref{fig1}.  It might also be objected that the eternal black holes that these spacetimes may describe are also unphysical idealizations.  Be that as it may,  they play an important role in the existing literature on  black hole thermodynamics \cite{HawkingBHT, HawkingPage} and also \cite{Maldacena2} in the theory of the AdS/CFT correspondence and we feel this is sufficient reason for our conjectures to be of interest.

We therefore study the free massless real scalar field (\ref{wave}) on the above flat 1+1 dimensional spacetime with vanishing boundary conditions on the mirror(s).  We show that there are classical solutions which are initially smooth with finite-energy tails -- namely certain initially right-moving classical solutions supported in regions IV, I and II -- which have an infinite value for the stress-energy tensor component, $T_{vv}$ (see below for notation) on the null line $t+x=0$.  (Below, because of the analogy with Schwarzschild and with Schwarzschild-AdS, we shall call this null line the \textit{B horizon}, ${\cal H}_B$ -- see the caption to Figure \ref{fig1}.)  Regarding the quantum case, we point out that many of these finite initial energy solutions (including all the ones we explicitly consider) also have finite norm in the appropriate one-particle Hilbert space (discussed below) and the quantum coherent states obtained by acting on the HHI vacuum (see below) with the quantum field smeared with the corresponding one-particle Hilbert space vectors correspond to the above classical solutions and, in particular, have the same singular $T_{vv}$ as those classical solutions, where $T_{vv}$ is now to be interpreted as the expectation value of the appropriately renormalized stress-energy tensor in those states.  Here, by `energy', we refer to the usual notion of energy that we would use in (1+1-dimensional) Minkowski space in the absence of any mirrors and by `initial energy' we mean that this is evaluated in Region IV $<$\ref{Note:energy}$>$.  (Later, we shall also consider an alternative, inequivalent, notion of [initial] energy for this 1+1-Minkowski mirror system  -- see below where we introduce the notion of \textit{re-signed Rindler energy}.)

\begin{figure} 
   \centering
    \includegraphics*[scale=0.5,bb=1.5in 3in 9in 9in]{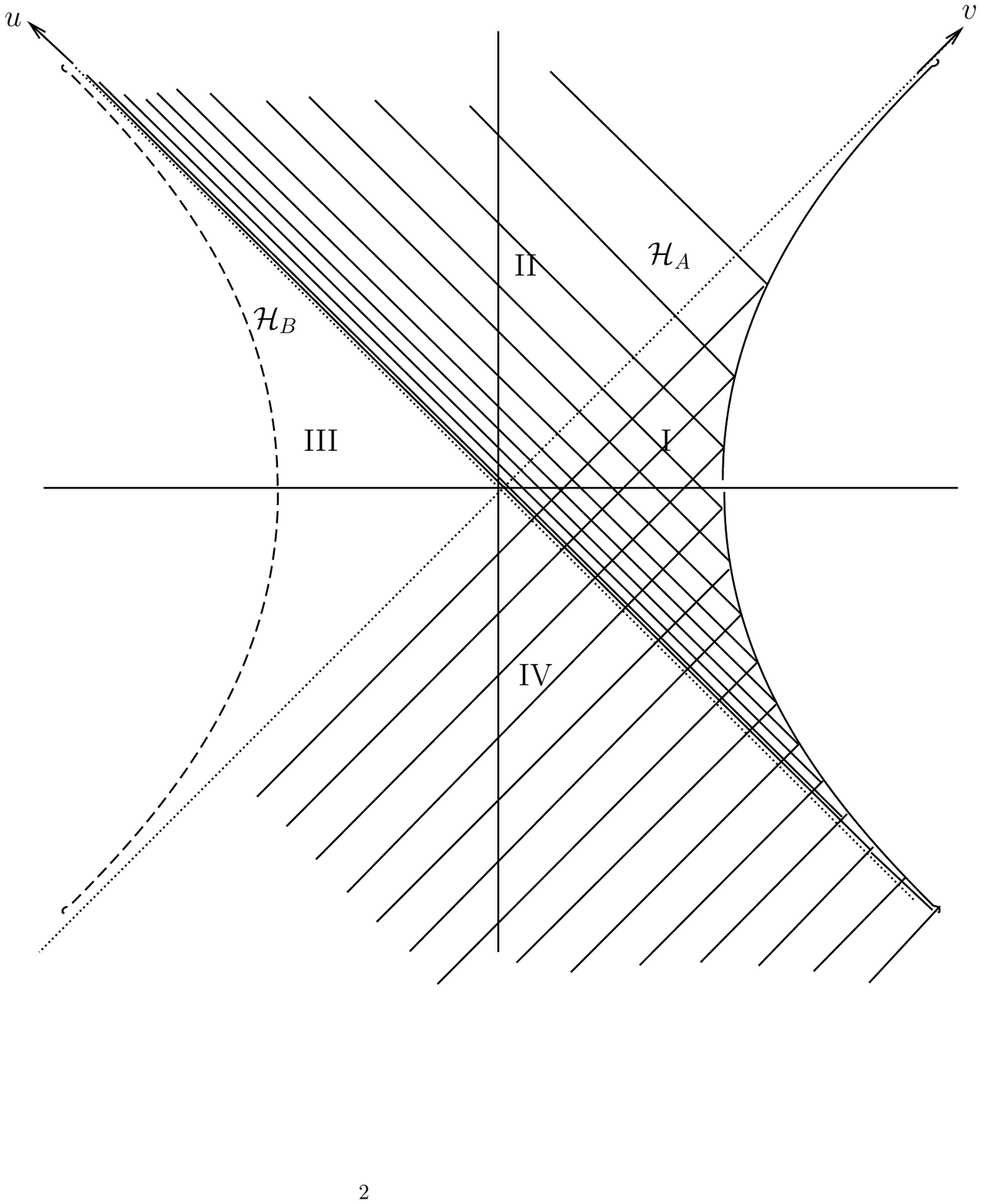}
   \caption{Lines of constant phase of (the restriction to Region IV of) an initially right-moving plane wave.  The wave reflects off the mirror in Region I and so do its lines of constant phase, piling up towards the horizon, ${\cal H}_B$.
\label{fig3}}
 \end{figure}

\section{A classical stress-energy singularity result for finite initial energy waves on the Minkowski with accelerated mirror(s) model}
\label{Sect:mirrorsing}

The origin of the infinity in $T_{vv}$ is simple:  An initially right-moving plane wave emanating from Region IV will have a given fixed phase on evenly spaced null lines as illustrated in Figure \ref{fig3}.  These will reflect off the mirror in Region I and pile up just to the future of the horizon ${\cal H}_B$.  To see this mathematically, let $t$ and $x$ be the usual time and space coordinates for our 1+1-dimensional Minkowski space and let $u=t-x$ and $v=t+x$ so our massless scalar field equation, 
\begin{equation}
\label{wave}
{\partial^2\phi\over\partial t^2} - {\partial^2\phi\over\partial x^2}=0 
\end{equation}
becomes 
\[
{\partial^2\phi\over\partial u\partial v}=0
\] 
with general solution 
\[ 
\phi(u, v)=f(u) + g(v),
\] 
a sum of right and left-going waves.   Assuming that both mirrors are present and are located on the two branches of the hyperbola $uv=-1$, our boundary condition becomes $\phi(u,v)=0$ at $uv=-1$, and $\phi(u,v)$ will clearly satisfy this if and only if $g(v)$ satisfies
\begin{equation}
\label{gfrelation}
g(v)=-f(-1/v). 
\end{equation}
It is to be understood here and below that in the two-mirror case we are only interested in the resulting $\phi(u,v)$ for $u$ and $v$ such that $uv>-1$ -- i.e.\ in the region between the two mirrors, while in the one-mirror case, this restriction only applies when $u$ is negative and $v$ is positive.  The picture in Figure \ref{fig3} then corresponds to taking $f(u)=e^{-i\omega u}$, for $u < 0$ and $0$ for $u \ge 0$ $<$\ref{Note:chiandonemirror}$>$, whereupon $g(v)=-e^{i\omega /v}$ for $0<v < 0$ and $0$ for $v\le 0$.   Here, and below, we adopt the convention that when we write complex classical solutions, $\phi(u, v)$, we intend their real part.

Of course, such plane-wave solutions will not have finite initial energy.  To see that there are finite initial energy solutions with the same pile-up property, and, in particular, the advertised singular $T_{vv}$, let us temporarily assume the left-hand mirror to be absent $<$\ref{Note:chiandonemirror}$>$ and take, for example, the solution,
$\phi_p^B(u,v)= f_p(u) + g_p(v)$, where 
\begin{equation}
\label{fp}
f_p(u)=\mathrm{const}(u^2+a^2)^{-p/2} e^{-i\omega u}
\end{equation}
where $a$ is a positive constant and $p$ a, possibly fractional, power and
\begin{equation}
\label{onegfrelation}
g_p(v) =- f_p(-1/v) \ \ \hbox{for} \ \ v>0 \ \hbox{and} \ \  g_p(v)=0 \ \ \hbox{for} \ \ v \le 0.
\end{equation}
This will have stress-energy tensor component, $T_{uu}=(d \mathrm{Re}[f(u)]/du)^2$ and thus will have initial energy $\int_{-\infty}^\infty (d\mathrm{Re}[f(u)]/du)^2 du$ which is easily seen to be finite provided $p > 1/2$ $<$\ref{Note:naturalQ}$>$.  (And its restriction to Region IV will have half this energy.) On the other hand, for $v>0$, $g_p(v)=-(1/v^2+a^2)^{-p/2}e^{i\omega/v}$ which is easily seen to have a stress-energy component, $T_{vv}$, which is singular at $v=0$ -- in the sense that it diverges as $v \rightarrow 0$ -- whenever $p<2$.  Thus, for each $p$ in the range $1/2 < p < 2$, we have a finite initial energy solution $<$\ref{Note:initial}$>$ for which the stress-energy tensor is singular on the horizon ${\cal H}_B$ (see Figure \ref{fig1}).  In particular, restricting to Regions IV and I, we have a finite initial energy solution consisting of an initially right-moving local solution in Region IV which, after reflecting on the mirror in Region I, becomes singular on the horizon, ${\cal H}_B$.  (Clearly there are many more finite-initial-energy solutions with the same property.)

We remark that, for $p$ in the range $3/2 < p < 2$, the total integrated energy of the reflected wave will be finite (even though there is a singularity in $T_{vv}$).  Thus for $p$ in this range, the total amount of work needed to be done on the right hand mirror is finite.   Also, since the stress-energy tensor at the mirror is finite, the force that needs to be exerted on the mirror to keep it on its trajectory is, at all times, finite $<$\ref{Note:realmirrors}$>$

\section{The quantum theory of the Minkowski two-mirror model}

\subsection{Construction of a Hartle-Hawking-Israel-like state}

For the quantum theory, assuming two mirrors, we can regard the region between them as the Wick rotation of the interior of a disk of radius 1 in a 1+1-dimensional flat Euclidean space and we can obtain (the time-ordered two-point function of) a preferred quantum state  -- the analogue of the Hartle-Hawking-Israel (\textit{HHI}) state \cite{Hartle-Hawking, Israel} -- by analytically continuing the Euclidean Green function for vanishing boundary conditions on the boundary of the disk.   This, in turn can be obtained rather easily by mapping the interior of the disk to the upper half complex plane with a M\"obius transformation, from the Green function on the upper half plane with vanishing boundary conditions on the real axis, and the latter is easily obtained by the method of images.   The result is that the (non-time-ordered) two-point function, $G(u_1,v_1; u_2,v_2)$, which can be taken to specify the theory (we assume for simplicity that the one-point function vanishes; if it doesn't this won't change our essential results) is given by
\[
\fl G(u_1,v_1; u_2,v_2)=
\]
\begin{equation}
\label{HHIG}
\fl -{1\over 4\pi}\log[(u_1-u_2-i\epsilon)(v_1-v_2-i\epsilon)]+{1\over 4\pi}\log[-((u_1-i\epsilon)(v_2+i\epsilon)+1)((v_1-i\epsilon)(u_2+i\epsilon)+1)]
\end{equation}
(where we take $\log$  to be real for positive argument and to have a branch cut on the negative real axis).
We should mention that, as always \cite{Wightman, Fulling-Ruijsenaars, KayNantes} with the $1+1$ dimensional massless scalar field, what is physically meaningful is not literally this two-point function but rather its $\partial^2/\partial u_1\partial u_2$, $\partial^2/\partial u_1\partial v_2$, $\partial^2/\partial v_1\partial u_2$ and $\partial^2/\partial v_1\partial v_2$ derivatives, so we may change the base of the logarithms or, e.g.\ replace the term $((u_1-i\epsilon)(v_2+i\epsilon)+1)$ by $((u_1-i\epsilon+1)/(v_2+i\epsilon))$ etc.\ without changing anything physical.

We wish to realize $G(u_1,v_1; u_2,v_2)$ as an expectation value
\begin{equation}
\label{GHilb}
G(u_1,v_1; u_2,v_2)=\langle \Omega| \hat\phi(u_1,v_1)\hat\phi(u_2,v_2)\Omega\rangle
\end{equation}
in a vacuum vector, $\Omega$, in a suitable Hilbert space, of a product of quantum fields, 
$\hat\phi(u,v)$, at different points.  As a seemingly good way to make mathematical sense of this -- loosely based on the developments in \cite{Kay-Wald} -- we begin by defining the space, $S=S_B+S_A$, of (real) classical solutions consisting of sums of  \textit{B-solutions} (whose $u$ derivatives have smooth compactly supported restrictions to the B horizon) of form 
\begin{equation}
\label{phiBdef}
\phi^B(u,v)=f(u)-f(-1/v) \  \hbox{for} \  f\in C_0^\infty(\mathbb R)
\end{equation}
and \textit{A-solutions} (whose $v$ derivatives have smooth compactly supported restrictions to the A horizon) of form 
\begin{equation}
\label{phiAdef}
\phi^A(u,v)=-g(-1/u)+g(v) \  \hbox{for} \ v\in C_0^\infty(\mathbb R).
\end{equation}
A useful remark is that $S_A$ and $S_B$ are far from being (symplectically) orthogonal.  Indeed, because of reflection at the mirrors, each solution in $S_B$ is equally well determined by its full restriction to \textit{either} horizon, although, while the restriction of its $u$-derivative to the B horizon will be smooth, neither the solution itself nor the restriction of its $v$-derivative to the A horizon will be smooth $<$\ref{Note:KWtechnicalities}$>$.  (And similarly vice versa.)

Rather than working with the mathematically problematic fields at a point, $\hat\phi(u,v)$, we work with `symplectically smeared' fields $\hat\phi(\phi^B)$ and $\hat\phi(\phi^A)$ for $\phi^A\in S_A$ and $\phi^B\in S_B$ formally related to the $\hat\phi(u,v)$ by
\[
\hat\phi(\phi^B)=2\int_{-\infty}^\infty \hat\phi(u,0){d\phi^B(u,0)\over du} du, \quad \hat\phi(\phi^A)=2\int_{-\infty}^\infty \hat\phi(0,v){d\phi^A(0,v)\over dv} dv
\]
where the integrals can be thought of as being, respectively, over the B and A horizons.  For general $\phi=\phi^A+\phi^B\in S$, we'll have $\hat\phi(\phi^A+\phi^B)=\hat\phi(\phi^A)+\hat\phi(\phi^B)$.
We next seek a \textit{one-particle structure} (see \cite{Kay-Wald} and the references there to earlier papers of the present author and/or \cite{KayPhD} in which this notion and notation was introduced) $(K, H)$,  for our space of solutions, $S$ -- that is a complex Hilbert space, $H$ (to be called our \textit{one particle Hilbert space}) and a real linear map, $K$, from $S$ to $H$ with dense range which  takes the natural symplectic form on any pair, $\phi_1, \phi_2$, of solutions in $S$, defined either (cf.\ our `useful remark' above) by the integral
$2\int_{-\infty}^\infty \phi_1(u)(d\phi_2(u)/du) du$ on the B horizon or by the integral $2\int_{-\infty}^\infty \phi_1(v)(d\phi_2(v)/dv) dv$ on the A horizon,
into twice the imaginary part of the inner-product of $K\phi_1$ and $K\phi_2$ in $H$.  We will then have that the double symplectic smearing,
$G(\phi_1, \phi_2)$, of $G(u_1,v_1; u_2,v_2)$ with a pair of classical solutions, $\phi_1(u_1,v_1), \phi_2(u_2,v_2) \in S$ is given by
\begin{equation}
\label{GHilbsmear}
G(\phi_1, \phi_2) \ =\langle K\phi_1|K\phi_2\rangle=\langle \Omega^F |\hat\phi^F(K\phi_1)\hat\phi^F(K\phi_2)\Omega^F\rangle
\end{equation}
where the inner product in the first equality is in the one-particle Hilbert space, while the second inner product is on the Fock space over $H$;  $\hat\phi^F(\psi)$, $\psi\in H$, denotes $-i(a^+(\psi)-a^+(\psi)^*)$ where $a^+(\psi)$ is the usual Fock-space creation operator for the one-particle vector $\psi$ and $a^+(\psi)^*$ its adjoint and $\Omega^F$ is the usual Fock-space vacuum vector.  We may then take (\ref{GHilbsmear}) to be a mathematically well-defined version of (\ref{GHilb}).

As a clue towards the correct definition of $(K,H)$, we compute the restriction to each of our A and B horizons of the distributional derivatives 
$\partial_{u_1}\partial_{u_2}G(u_1,v_1; u_2,v_2)$ and $\partial_{v_1}\partial_{v_2}G(u_1,v_1; u_2,v_2)$ and find, easily, from (\ref{HHIG}) that these have the universal (cf.\ \cite{Kay-Wald}) forms
\begin{equation} 
\label{universalB}
\partial_{u_1}\partial_{u_2}G(u_1,0; u_2,0)=-{1\over 4\pi (u_1-u_2-i\epsilon)^2}
\end{equation}
\begin{equation}
\label{universalA}
\partial_{v_1}\partial_{v_2}G(0,v_1; 0,v_2)=-{1\over 4\pi (v_1-v_2-i\epsilon)^2}.
\end{equation}

It is straightforward to infer from (\ref{universalB}) and (\ref{universalA}) that (cf.\ the derivation of Eq. (4.19) in \cite{Kay-Wald}), for a pair, $\phi^B_1, \phi^B_2$ in $S_B$, 
\begin{equation}
\label{KBKB}
\langle K\phi^B_1|K\phi^B_2\rangle=-\lim_{\epsilon\rightarrow 0}\left ({1\over \pi}\int_{-\infty}^\infty\int_{-\infty}^\infty {f_1(u_1)f_2(u_2)\over (u_1-u_2-i\epsilon)^2}du_1du_2\right )
\end{equation}
where $f_1$ and $f_2$ are related to $\phi^B_1$ and $\phi^B_2$ as in (\ref{phiBdef}) and, similarly, for a pair, $\phi^A_1, \phi^A_2$ in $S_A$,
\begin{equation}
\label{KAKA}
\langle K\phi^A_1|K\phi^A_2\rangle=-\lim_{\epsilon\rightarrow 0}\left ({1\over \pi}\int_{-\infty}^\infty\int_{-\infty}^\infty {g_1(v_1)g_2(v_2)\over (v_1-v_2-i\epsilon)^2}dv_1dv_2\right )
\end{equation}
where $g_1$ and $g_2$ are related to $\phi^A_1$ and $\phi^A_2$ as in (\ref{phiAdef}).

To complete the specification of $(K,H)$ we need also to know $\langle K\phi^B_1|K\phi^A_2\rangle$.  For typical spacetimes with bifurcate Killing horizons \cite{Kay-Wald}  (e.g.\ Schwarzschild) this is difficult to compute.  However, for our 1+1 dimensional massless scalar field in Minkowski space with our mirrors, we must have, in view of our above useful remark, both of the equalities:
\begin{equation}
\label{KBKA1}
\langle K\phi^B_1|K\phi^A_2\rangle=\lim_{\epsilon\rightarrow 0}\left({1\over \pi}\int_{-\infty}^\infty\int_{-\infty}^\infty {f_1(u_1)g_2(-1/u_2)\over (u_1-u_2-i\epsilon)^2}du_1du_2\right)
\end{equation}
and
\begin{equation}
\label{KBKA2}
\langle K\phi^B_1|K\phi^A_2\rangle=\lim_{\epsilon\rightarrow 0}\left({1\over \pi}\int_{-\infty}^\infty\int_{-\infty}^\infty {f_1(-1/v_1)g_2(v_2)\over (v_1-v_2-i\epsilon)^2}dv_1dv_2\right).
\end{equation}
It is easy to check directly, by making the substitutions $u_1=-1/v_1$, $u_2=-1/v_2$ in (\ref{KBKA2}), that the right-hand sides of (\ref{KBKA1}) and (\ref{KBKA2}) are indeed equal: The integral over $v_1$ and $v_2$ in (\ref{KBKA2}) transforms to an integral over $u_1$ and $u_2$ which is the same as that in (\ref{KBKA1}) except that the $\epsilon$ in the latter is replaced by $\epsilon u_1u_2$  but this of course gives the same result in the limit.  One can, alternatively, infer (\ref{KBKA1})/(\ref{KBKA2}) directly from (\ref{HHIG}).  By the way, the same integral substitution enables one easily to check that the right-hand side of  (\ref{KBKB}) remains unchanged if one replaces $f_1(u_1)$ by $f_1(-1/u_1)$ and 
$f_2(u_2)$ by $f_2(-1/u_2)$ (and similarly for (\ref{KAKA})).  Using this, one sees that
the Cauchy Schwartz inequality, $\langle K\phi^B_1|K\phi^A_2\rangle^2\le \langle K\phi^B_1|K\phi^B_1\rangle\langle K\phi^A_2|K\phi^A_2\rangle$, holds and thus that our putative inner-product on $H$ really is an inner-product and so $H$ really is a Hilbert space and therefore our two-point function $G$ satisfies the necessary positivity requirements to be the two-point function of a genuine state.    (An alternative demonstration of the latter could proceed by checking that a suitable version of reflection positivity -- see e.g.\ \cite{Jaffe-Ritter} and references therein -- holds for our Euclidean Green function.)

With $(K,H)$ defined as in (\ref{KBKB}), (\ref{KAKA}), (\ref{KBKA1})/(\ref{KBKA2}) it is straightforward to check that (\ref{GHilbsmear}) holds.

The two-point function (\ref{HHIG})/(\ref{GHilb})/(\ref{GHilbsmear}) is well-behaved in many respects.  In particular, the renormalized expectation values, $\langle T_{uu}^{\mathrm{ren}}\rangle_{\mathrm{HHI}}$, 
$\langle T_{vv}^{\mathrm{ren}}\rangle_{\mathrm{HHI}}$
in our HHI state of the components  $T_{uu}$ and $T_{vv}$ of the renormalized stress-energy tensor are everywhere zero!  [The trace term, (4 or -4 times) $T_{uv}$ (depending on one's choice of signature), vanishes identically classically and, since there is no trace-anomaly (see e.g.\ \cite{Birrell-Davies}) for the massless free scalar field in a (locally) flat spacetime, also quantum mechanically.]

In fact we have (cf.\ \cite{KayCasimir})
\begin{equation}
\label{rnTuu}
\fl\langle T_{uu}^{\mathrm{ren}}(u,v)\rangle_{\mathrm{HHI}}=\lim_{(u_1-u)^2+(v_1-v)^2+(u_2-u)^2-(v_2-v)^2\rightarrow 0} {\partial\over\partial u_1}{\partial\over \partial u_2} 
\left (G(u_1,v_1; u_2,v_2) - G_0(u_1,v_1; u_2,v_2)\right),
\end{equation}
\begin{equation}
\fl\label{rnTvv}
\langle T_{vv}^{\mathrm{ren}}(u,v)\rangle_{\mathrm{HHI}}=\lim_{(u_1-u)^2+(v_1-v)^2+(u_2-u)^2-(v_2-v)^2\rightarrow 0} {\partial\over\partial v_1}{\partial\over \partial v_2} 
\left (G(u_1,v_1; u_2,v_2) - G_0(u_1,v_1; u_2,v_2)\right)
\end{equation}
where $G_0$ denotes the two-point function in the usual Minkowski vacuum state (i.e.\ in the absence of our mirrors).  One easily sees that both of these vanish, on noticing that $G_0$ is equal to the first term of $G$ in (\ref{HHIG}) and therefore cancels it.

The quantity $\langle\partial_u\hat\phi\partial_v\hat \phi\rangle_{\mathrm{HHI}}^{\mathrm{ren}}$ (which is not a component of the stress-energy tensor) can be calculated similarly and one easily finds 
\begin{equation}
\label{mirrordiverge}
\langle\partial_u\hat\phi\partial_v\hat \phi)\rangle_{\mathrm{HHI}}^{\mathrm{ren}}=
{1\over 4\pi}{1\over (uv+1)^2}
\end{equation}
which is also smooth in the interior of the spacetime and, in particular, finite (with value $1/4\pi$) on the horizons, albeit it diverges as the mirrors are approached.

\subsection{The quantum version of our stress-energy singularity result our Minkowski two-mirror model}
\label{Sect:quantumversion}

However, the situation for certain non-vacuum states is less well-behaved:   Consider coherent states of the form $e^{-i\hat\phi(\phi)}\Omega^F=\exp(-\langle K\phi|K\phi\rangle)e^{-a^+(K\phi)}\Omega^F$ where $\phi$ is a suitable classical solution.  
Here, we intend, by a `suitable' solution, not only a solution belonging to $S=S_B+S_A$ as defined above, but  also any other solution, $\phi$ which arises as $\phi^B$+$\phi^A$, where $\phi^B$ and $\phi^A$ are of the form of (\ref{phiBdef}) and (\ref{phiAdef}) provided only that the right-hand sides of $\langle K\phi^B|K\phi^B\rangle$ and $\langle K\phi^A|K\phi^A\rangle$, defined as in (\ref{KBKB}) and (\ref{KAKA}), are finite.  In particular, one may check that any of our classical solutions, ${\phi'}_p^B$ -- see Endnote $<$\ref{Note:chiandonemirror}$>$ -- where $p>1/2$, is suitable in this sense $<$\ref{Note:checksuitable}$>$.  For any such coherent state, it is easy to see that the expectation value of the renormalized stress-energy tensor is equal to the classical stress-energy tensor of the classical solution, ${\phi'}_p^B$, and the expectation value of the total renormalized `initial' energy on the B horizon is equal to the classical total `initial' energy on the B horizon $<$\ref{Note:easycoherent}$>$.  

Thus we conclude that (provided, in the quantum case, we interpret `value' to mean `expectation value' and both `initial' energy and stress-energy tensor are assumed to be renormalized) in both the classical and the quantum theory, there are states for which the value of the `initial' total energy on the B horizon is finite and the `initial' stress-energy tensor (i.e.\ the $uu$ component of the stress-energy tensor) is everywhere finite but for which the value of the ($vv$ component of the) stress-energy tensor diverges as we approach the B horizon.   (Note that we have used the word `initial' here, and in the last paragraph of Endnote $<$\ref{Note:chiandonemirror}$>$,  in a different sense from its usual meaning and, for this reason, have put the word between quotes.   Below, we return to using the word with its more usual meaning of `at early times'.)

\section{Conjectured generalization to other enclosed horizons} 
\label{Sect:conjecture}

We expect that there will be a similar (classical and quantum) singularity in the stress-energy tensor for initial finite energy, initially, say, smooth, solutions for other enclosed horizons when their picture resembles Figure \ref{fig1}.   More precisely we expect this when the enclosure is \textit{asymptotically null} and the suppressed dimensions in Figure \ref{fig1} are \textit{compact}.   We shall tacitly assume throughout the rest of the paper that when we refer to a spacetime with enclosed horizons, it satisfies both of these conditions.  Both these conditions hold in our Schwarzschild and Schwarzschild-AdS examples.  

The reason for the asymptotically null condition is that this property in the 1+1 Minkowski with mirror(s) model is clearly needed for the pile-up of plane waves near the horizon, ${\cal H}_B$, discussed at the beginning of Section \ref{Sect:mirrorsing} and for the pile-up
of the solutions with finite energy tails,  $\phi_p^B$, near ${\cal H}_B$.   We remark (see Section \ref{Sect:AdSabsence}) that an example of an enclosed horizon which is not asymptotically null is provided by plain AdS.  The enclosure (i.e.\ conformal boundary) in that case is timelike, rather than null.   The difference is reflected in the fact that in the conformal compactification, the horizon actually reaches the conformal boundary at a corner which forms part of that boundary --  in contrast to Schwarzschild-AdS and other cases with asymptotically null enclosures, where such a corner is missing.

The reason for assuming compactness of the extra dimensions is illustrated e.g.\ by the case of our 1+1 Minkowski mirror system producted with a flat 2 torus.
By Fourier analysis in the extra dimensions, massless waves in this case will obviously decompose as discrete sums of (now all-but-one massive) waves on 1+1 Minkowski for which, again one expects something like the pile-up in the 1+1 massless case and for which one of course has the same pile-up as in the 1+1 massless case in the toroidally symmetric sector.

\begin{figure} 
   \centering
    \includegraphics*[scale=0.5,bb=1.5in 3in 9in 9in]{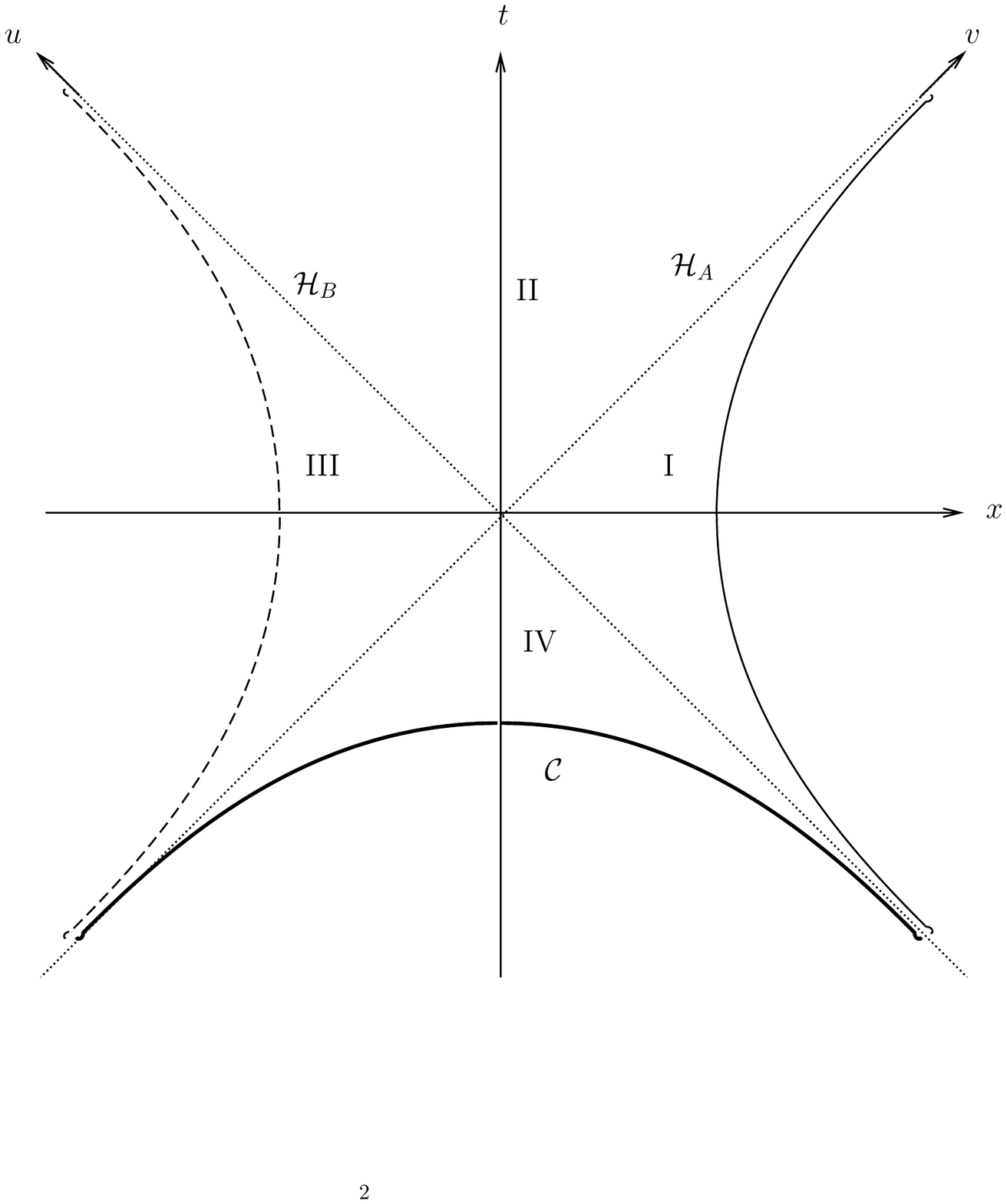}
   \caption{The surface $\cal C$, $t^2-x^2=\mathrm{const}^2$, $t<0$.  In the presence of both mirrors, the Cauchy problem for the 1+1 wave equation will be well posed for data on $\cal C$. 
\label{fig4}}
 \end{figure}

Once we go beyond our Minkowski-mirror system and consider e.g.\ the Schwarzschild example, (now, Kruskal) $t$ translations are no longer (local) isometries so we have to be careful what we mean by the total initial energy of a perturbation.  However, we could consider solutions specified by their Cauchy data on a suitable initial surface, say (see Figure \ref{fig4})  $t^2-x^2=\mathrm{const}^2$, $t<0$, in Region IV and regard them as having finite energy if the integral of $T_{uu}$ over the negative-$u$ half of the B horizon and of $T_{vv}$ over the negative-$v$ half of the A horizon are each finite, where $u$ and $v$ now denote the affine parameter along the horizon generators (i.e.\ the usual Kruskal null-coordinates).  

Alternatively (and inequivalently) we could say that a solution's (initial -- but see next remark) energy is finite if the restriction of the solution to Region I has finite (positive) energy with respect to Schwarzschild time (equivalently with respect to $\log u$ on the negative-$u$ half of the B-horizon, missing out a, here, inessential factor of the surface gravity)  and their restriction to the left wedge has finite (negative) energy with respect to the same Schwarzschild time-evolution -- extended in the usual way to the action of the Schwarzschild isometry group to all four wedge regions.  So we seem to get a useful notion of energy for the Schwarzschild interpretation of Figure 1 (with box walls in both Region I and III) by defining it to be its energy with respect to Schwarzschild time in Region I \textit{minus} its energy with respect to Schwarzschild time in Region III.  Thanks to the minus sign, this will be positive for all solutions.  We remark that because the Schwarzschild isometry group maps the box walls to themselves, the Schwarzschild energy is conserved (separately) in Region I and Region III so this energy is an attribute of a solution, and need not be thought of just as an `initial energy in Region IV'.  We shall call it the solution's \textit{re-signed Schwarzschild energy}.   

Defining energy analogously to the latter way in our 1+1-dimensional Minkowski system with two mirrors (where we might call the analogous energy-notion finite \textit{re-signed Rindler} energy -- cf.\ \cite{Rindler} and Endnote $<$\ref{Note:Rindlerfootnote}$>$) it is easy to see that ${\phi'}_p^B$ (see Endnote $<$\ref{Note:chiandonemirror}$>$) will still have finite energy in this latter sense precisely when the integral of $uT_{uu}$ over the negative-$u$ half of the B horizon is finite and that this will hold  provided $p>1$ (and similar statements hold of course, for the A-horizon with $u$ replaced by $v$.)  So we still get a similar stress-energy singularity result for $1<p<2$.  We conjecture that, quite generally, for linear scalar Bose fields, as long as the suppressed dimensions in Figure \ref{fig1} are compact -- as is the case in Schwarzschild or Schwarzschild-AdS or as would be the case, e.g., for the product of our 1+1 dimensional Minkowski mirror example with a flat 2-torus -- there will be classical solutions  with finite re-signed Schwarzschild energy  (and its obvious counterpart in Schwarzschild-AdS, which we shall give the same name, and either finite re-signed Rindler energy or finite initial Minkowski energy in Region IV in the Minkowski times torus case etc.) for which the stress-energy tensor is singular on the A and B horizons $<$\ref{Note:characteristicversion}$>$.  And we expect a similar result for the electromagnetic and linearized gravitational fields.   

We remark that in the cases, e.g., of Schwarzschild in a box or Schwarzschild-AdS, the finite re-signed Schwarzschild energy perturbations which are, on this conjecture, responsible for the singularity in the stress-energy tensor, come, when tracked back in time, from the (usual) past (Schwarzschild spacelike) spacetime singularity.  But this doesn't in any way detract from the significance of our results/conjectures.

\section{Comparison with the stress-energy singularity on the Cauchy horizon of Reissner Nordstr\"om}
\label{Sect:RNcomparison}

It is interesting to compare and contrast this conjectured stress-energy singularity result with the well-known and well-established stress-energy singularity result for the Cauchy horizon of the (non-extremal) Reissner-Nordstr\"om solution \cite{Simpson-Penrose, Hawking-Ellis, Chandra-Hartle}.  Just as an observer crossing the Cauchy horizon of Reissner-Nordstr\"om would \cite{Hawking-Ellis} ``see the whole history of one of the asymptotically flat regions in a finite time'' so an observer passing from Region IV to Region I in Figure \ref{fig1} would  see an infinite amount of the history of Region IV in a finite amount of time $<$\ref{Note:history}$>$, albeit this would be seen just \textit{after} crossing the horizon and in reverse time-order.  

In both cases, these similar facts about the classical geometry might lead one to suspect  that there are, in some suitable sense, small initial perturbations for which the stress-energy tensor is singular on the horizon as we have demonstrated here in our 1+1 dimensional Minkowski case (and as we have now conjectured for other cases) of enclosed horizons and as was verified in \cite{Simpson-Penrose, Chandra-Hartle} for Cauchy horizons.  However, we should draw attention to two significant differences:  Firstly, in the Reissner-Nordstr\"om Cauchy-horizon case, it was shown \cite{Hiscock} that the renormalized quantum stress-energy tensor of a linear scalar field in the HHI state diverges as one approaches the Cauchy horizon and this seemed in accordance with a general expectation that the classical stress-energy singularity goes hand-in-hand with a pathology in the relevant preferred quantum state.  Here, for our 1+1-Minkowski two-mirror model, we have seen that the appropriate HHI state has a renormalized stress-energy tensor which is finite (in fact zero) on the relevant (Killing) horizon.  However, we saw that there are quantum coherent states for which it diverges.  So, to summarize: In our 1+1-Minkowski two-mirror system, it would seem that the singularity in the classical stress-energy tensor goes hand-in-hand with a quantum singularity, but the relevant preferred quantum state (our HHI state) is, itself, free from any obvious pathology.  

Secondly, and, arguably, more troublingly, the precise nature of the `small initial data' which lead to a singular stress-energy tensor at the relevant horizons is different.   For the Cauchy horizon of  Reissner-Nordstr\"om, this stress-energy singularity holds for smooth compactly supported initial Cauchy data on a suitable initial surface $<$\ref{Note:any}$>$.   For our enclosed horizons here,  our stress-energy singularity result in our 1+1-Minkowski model with (one or two) mirrors (and similarly our conjecture in other cases) held, it is true, for finite initial-energy, initially smooth, solutions,  but it was essential that $f(u)$ \textit{had an infinitely extended tail} for large negative $u$; a classical solution in $S_B$, i.e.\ for which $f$ is \textit{compactly supported}, will have a \textit{finite} stress-energy tensor on the horizon ${\cal H}_B$.   To help in the subsequent discussion, we shall say that a stress-energy singularity result is \textit{gold-plated} if it holds for smooth compactly supported data on a suitable initial surface; in this sense, the Reissner Nordstr\"om result is gold-plated, our stress-energy singularity result/conjecture for enclosed horizons is not.  

\section{Towards a stress-energy `almost-singularity' result with compactly supported initial data}
\label{Sect:silver}

Partly motivated by the above comparison, we next discuss the prospects for having something similar to a gold-plated stress-energy singularity result for our 1+1-Minkowski system with (one or) two mirrors.  More precisely we shall ask if it is possible to have something similar to a singular stress-energy tensor on our horizons for (small) smooth \textit{compactly supported} initial data on some suitable initial surface.  (Another motivation for this investigation is indicated in Endnote $<$\ref{Note:naturalQ}$>$.)

Consider the initial surface, $t^2-x^2=\mathrm{const}^2$, $t<0$, sketched in Figure \ref{fig4}.  While this is not quite a Cauchy surface, in view of our boundary conditions on our (here, we assume, two) mirrors, the Cauchy problem for our 1+1 dimensional massless scalar wave equation with our two mirrors will clearly be well posed for smooth compactly supported data on it.  One can easily convince oneself that -- in some fixed Lorentz frame -- by choosing the support of its Cauchy data on this surface to be located at sufficiently large positive $x$ (and consequently large negative $t$) an arbitrarily small right-going solution with data supported in a small region of this initial surface with everywhere arbitrarily small $T_{uu}$, can be chosen to have a $T_{vv}$ as large as we like along some null line parallel to and just to the right of ${\cal H}_B$.  And of course a similar argument can be made that such data on our initial surface supported in a small region located at large negative $x$ (and therefore again large negative $t$) will give rise to a large $T_{uu}$ near ${\cal H}_A$.  

In fact, and to make this more precise, let us take a `small' solution here to mean a solution which has a small re-signed Rindler energy.   Then, by taking compactly supported Cauchy data for a right-going solution on our above initial surface, which is arbitrarily small in this sense, and applying a sufficiently large negative boost so as to map the support to a region of our initial surface at sufficiently large positive $x$, the solution determined by it will $<$\ref{Note:boost}$>$, due to reflection on the mirror in Region I, clearly have a $T_{uu}$ as small as we like on the initial surface while having a $T_{vv}$ as large as we like along some null line parallel to and just to the right of ${\cal H}_B$.   We remark that the re-signed Rindler energy of the thus boosted data will have the same (arbitrarily small) value as before the boost.  (And, by the way, its Minkowski energy after the boost -- were we to prefer that as a measure of smallness -- would be smaller!)   

One might think that this result, involving arbitrarily large stress-energy components for arbitrarily small initial data with arbitrarily small initial stress-energy tensor components, would be a reasonable substitute for a gold-plated result.   However, there is an important and somewhat subtle difficulty:  `large' is not the same as `singular', and, given such a solution, there will clearly be another Lorentz frame (where the support of the data on our initial surface has been boosted back to lie, say, around $x=0$) in which $T_{vv}$ is everywhere not large. {\it In the case of two mirrors} we can, however, overcome this difficulty as follows:  Choose compactly supported initial data on our initial surface which is the sum of (again arbitrarily small) compactly supported data for a right-going solution and of (arbitrarily small) compactly supported data for a left-going solution.   Then by applying a sufficiently large negative boost to the right-going data -- so as to map it to a region of our initial surface with large positive $x$ -- and a sufficiently large positive boost to the left-going data -- so as to map it to a region of our initial surface with sufficiently large negative $x$ -- we may clearly make $T_{vv}$ as large as we like near ${\cal H}_B$ and simultaneously make $T_{uu}$ as large as we like near ${\cal H}_A$ while $T_{uu}$ and $T_{vv}$ are as small as we like on our initial surface.  Now, no matter what Lorentz frame we transform to, at least one of $T_{vv}$ near ${\cal H}_B$ and  $T_{uu}$ near ${\cal H}_A$ will still clearly be large somewhere (in fact there will be a null line on which one or other of these quantities gets even larger)!   Moreoever (I am very grateful to Chris Fewster who, after I showed him my argument up to this point, suggested that something along the lines which follow might hold thanks to `interference terms') near the bifurcation point -- i.e.\ the point where ${\cal H}_A$ and ${\cal H}_B$ intersect -- there will be a subregion of Region II where the solution with such initial data will consist of a sum of non-vanishing right-going and left-going waves and -- by making our above negative and positive boosts sufficiently big -- the (now Lorentz-invariant!) quantity $T_{uu}T_{vv}$ can clearly be made as large as we like somewhere in that region (while there is still a frame in which both $T_{uu}$ and $T_{vv}$ are small on our initial surface) $<$\ref{Note:infinitescalar}$>$! 

We remark that (still in the case of two mirrors) an obvious quantum version of this \textit{silver-plated} stress-energy \textit{almost-singularity} result will go through along similar lines to the quantum version of our result on the singularity in the classical stress-energy tensor for incoming waves with suitable finite-energy tails as explained in Section \ref{Sect:quantumversion}.

We feel that, taken together, all these results are a reasonable substitute for a gold-plated stress-energy singularity result:  We might say we have, instead, a \textit{silver-plated} stress-energy \textit{almost-singularity} result --  i.e.\ the existence of arbitrarily small compactly supported Cauchy data on our initial surface with (in some Lorentz frame) arbitrarily small initial values for $T_{uu}$ and $T_{vv}$ which lead, in every  Lorentz frame, to one or other of $T_{uu}$ as large as we like somewhere near ${\cal H}_A$ and $T_{vv}$ as large as we like near ${\cal H}_B$ as well as to a (Lorentz-boost-invariant) $T_{uu}T_{vv}$ ($=T_{ab}T^{ab}/4$) as large as we like somewhere near the bifurcation point $<$\ref{Note:straightalternative}$>$  in Region II.

We end this subsection we make two further remarks:  First, that, in the case where there is no image mirror at the dashed hyperbola in Figure \ref{fig1}, we still have our original stress-energy singularity result for incoming waves with finite energy tails and we still have arbitrarily small compactly supported Cauchy data on our initial surface with (in some Lorentz frame) arbitrarily small initial values for $T_{uu}$ and arbitrarily large values of $T_{vv}$ near ${\cal H}_B$ -- but there is the objection, which we discussed above, that this is a Lorentz-frame dependent result.  Second, as we explain in Endnote $<$\ref{Note:realmirrors}$>$ we expect all the above results to be robust enough to survive when we modify the mirror trajectory (/trajectories) so as e.g.\ to be non-inertial for only a finite (but arbitrarily large) interval of proper time, and also to survive, for sufficiently large flat mirrors transversal to the direction of their motion, in 1+3 dimensions.

\subsection{Conjectured generalization to other enclosed horizons}
\label{Sect:silverconjecture}

We conjecture that similar silver-plated stress-energy almost-singularity results will hold for other (now 1+3-dimensional) spacetimes with enclosed horizons  including our 1+1-Minkowski two-mirror model times a flat two-torus, as well as Schwarzschild in a box (with a box wall in the right wedge and an image box wall in the left wedge) and Schwarzschild-AdS when pictured as in Figure 1 -- and, in particular, that we will still have arbitrarily large (boost-invariant/Schwarzschild-isometry-invariant etc.) $T_{uu}T_{vv}$ somewhere in Region II near the bifurcation (now) surface, where $T_{uu}$ and $T_{vv}$ are now the $uu$ and $vv$ components of the now 1+3 dimensional stress-energy tensor for (linear scalar, electromagnetic or linearized gravitational) fields of interest.   For Schwarzschild in a box and for maximally extended Schwarzschild-AdS, of course we have to choose our initial surface to lie to the future of the past spacetime singularity.  (E.g.\ in the case of Kruskal we can choose the $t<0$ branch of a hyperboloid $t^2-x^2= {\mathrm{const}}^2$, $t$ and $x$ being Kruskal coordinates, where the constant is chosen so that it lies to the future of the past spacetime singularity.)

\section{Discussion}
\label{Sect:Discussion}

\subsection{The argument for the instability of enclosed horizons}
\label{Sect:instabilityargument}

As is well-known, the stress-energy singularity result for the Reissner Nordstr\"om Cauchy horizon is believed to indicate the instability of that Cauchy horizon, in the sense of a big change in the spacetime geometry near it, once Newton's constant is switched on and  matter and gravity perturbations are allowed for.   Presumably it becomes, instead, a curvature singularity.   (See again \cite{Simpson-Penrose, Hawking-Ellis, Chandra-Hartle} and see also \cite{Dafermos1, Dafermos2}.)   In view of our original stress-energy singularity conjecture (for initially finite energy solutions with tails) and bolstered now by our above silver-plated stress-energy almost-singularity conjecture (which involves arbitrarily small smooth compactly supported initial data) we think it's reasonable to assume that, similarly,  for $1+3$-dimensional models with (asymptotically null and with compact suppressed dimensions in Figure \ref{fig1}) enclosed horizons resembling Figure \ref{fig1}, at least in the case when there is an image box wall/mirror, as soon as small matter and gravity perturbations are allowed for and the coupling to gravity is switched on, there will be a big change in the spacetime geometry around our horizons ${\cal H}_A$, ${\cal H}_B$.    Indeed, were that not to be the case, then, for example, in our 1+1-Minkowski two-mirror model times a flat two-torus, by Einstein's equations and, for simplicity assuming the trace of the stress-energy tensor to vanish  -- as would be the case for a conformally coupled massless field -- and considering, say, arbitrarily small initial perturbations of such a field which are toroidally symmetric, we would have, near the bifurcation surface, that (setting $8\pi G/c^4=1$) the scalar curvature invariant $R_{ab}R^{ab}$ would equal $T_{ab}T^{ab} = 4T_{uu}T_{vv}$ which, as we argued above, would be arbitrarily large somewhere near the bifurcation surface, in contradiction with the local flatness of the geometry there.   Again, in view of this and the remaining statements in our stress-energy singularity and almost-singularity conjectures, it seems reasonable to expect that the horizons become curvature singularities.  It is not difficult to see that, with these same conjectures, a similar conclusion can be similarly argued for for Schwarzschild in a box (with an image box) and for Schwarzschild-AdS (say for spherically symmetric initial perturbations).  

As we saw above, in our 1+1 dimensional model, the singularity in the classical stress-energy tensor for incoming waves with suitable finite-energy tails is clearly present whether or not we assume the presence of an image box wall/mirror in Region III of Figure \ref{fig1}.   However, as we already emphasized, our above silver-plated stress-energy almost-singularity result requires the presence of both mirrors.   Thus it is maybe less clear whether or in just what sense, say, the product of our 1+1-Minkowski single-mirror spacetime with a 2-torus would be unstable due to initial perturbations with compact support.  A similar remark would also apply to a Schwarzschild black hole in a box if we model it without an image box wall in the left wedge.   But of course, in Schwarzschild-AdS, there are mirror-like (conformal) boundaries in both the left and right wedges so all the above conjectures, assumptions and arguments should apply to that case.   We also note that, when there is only a mirror in Region I and no image mirror in Region III, the quantum theory of our 1+1-Minkowski mirror system will be quite different from that discussed above when both mirrors are present.  Further work on both the classical and quantum theory of this latter one-mirror (/one box) case is in progress \cite{Kay-Lupo}.   (See also our further mention of this work at the end of Section \ref{Sect:blackhole}.)

\subsection{Quantum implications and the correct spacetime description of black holes in equilibrium in spherical boxes}
\label{Sect:blackhole}

Assuming our conjectures in Sections \ref{Sect:conjecture} and \ref{Sect:silverconjecture} and our assumptions in Section \ref{Sect:instabilityargument} hold, how should we picture a classical spherical black hole in equilibrium (with its Hawking radiation atmosphere) in a (stationary spherical) box?  If, as we have argued -- at least in the case there is an image box wall in Region III -- the picture in Figure \ref{fig1} is unstable $<$\ref{Note:unstable}$>$, it is tempting to assume the correct picture (at least if we demand time-reversal symmetry -- see next paragraph) is something like Figure \ref{fig2}  (Regions II, III and IV, being separated from Region I by horizons which have now become curvature singularities, dropping out of the story.   Of course, from the standpoint of Region III, it would be Regions I, II and IV which dropped out the story etc.)  

As far as the quantum theory is concerned, we expect that, in a quantum field theory in curved spacetime approximation, there will be an HHI state on the region of the Kruskal spacetime between the box wall and the image box wall (assuming it to be present) as depicted in Figure \ref{fig2}.  But when quantum gravity is switched on, we expect there will be arbitrarily small perturbations of this state in which expected curvatures get extremely large near the horizons.  As a result, it seems reasonable to expect -- and we shall assume this from now on --  that in full quantum gravity there will be no counterpart to the HHI state on the full region of Kruskal between the box wall and the image box wall and, instead, the state describing a spherical black hole in equilibrium in a box will have an approximate description in terms of a classical spacetime resembling Figure \ref{fig2}.

Above we have attempted to discuss the classical and quantum situations separately but, on the above assumption, presumably there is no fully satisfactory fully classical description and the curvature singularity where the horizons bounding Region I used to be is really an inadequate classical representation of a situation (say in the region between the horizon and the inner hyperbola in Figure \ref{fig2} -- see Endnote $<$\ref{Note:Deltar}$>$) that really requires a quantum gravitational description and cannot be described in classical terms.  All in all, we end up with a picture very similar to that presupposed in the brick-wall model \cite{tHooft} previously argued for by 't Hooft in 1985 on different grounds.   

Similarly, and again assuming our conjectures and assumptions to hold, we assume that (time-reversal symmetric) quantum Schwarzschild-AdS would look like Figure \ref{fig2} and not (as Maldacena suggested in \cite{Maldacena2}) like Figure \ref{fig1}.  

Of course, another possibility, which is, however, not time-reversal-symmetric, would be to assume that a black hole in equilibrium in a box should be pictured as the union of Regions I and II only of Figure \ref{fig1} (with the usual future Schwarzschild spacelike singularity in Region II).  Now (cf.\ our above remark about being able to adopt the view that our initial matter perturbations come from infinite past Schwarzschild times in the exterior Schwarzschild region I) we would still expect a quantum region in the stress-energy tensor all along ${\cal H}_B$.  But that's OK, since this is now at the edge of our spacetime and we again suppose that what it is telling us is that, near this edge, the classical description breaks down.   Common sense might seem to suggest that if we were to capture a physical black hole (formed originally by stellar collapse) and put it in a box and let it come into equilibrium (with its Hawking radiation) this might be what we would get.  However, it is difficult to imagine that a stable equilibrium state would not be time-reversal-symmetric.  Whereas, on noting, e.g.\ that the null boundaries of Region I would now have a different nature -- the right part of ${\cal H}_B$ being a quantum region, while the right part of ${\cal H}_A$ remaining an ordinary classical (future) horizon, we see that, even restricting to Region I, there would be time-reversal asymmetry.  While it may seem to defy common sense, we feel that it is reasonable to entertain the possibility that, as the system equilibrates, Region II fades from the picture and we are left with a picture like Figure \ref{fig2} (i.e.\ the picture presupposed in the 't Hooft brick wall model).  Moreover, e.g.\ but not only, in the Schwarzschild-AdS situation, if what we are interested in is the correct Minkowski-signature description of the state defined by the Euclidean path integral, then we would expect this to be time-reversal-symmetric and therefore (on our conjectures) this correct Minkowski-signature description would be like Figure \ref{fig2}.   

All this of course resonates with the long-lasting Hawking-Penrose debate \cite{Hawking-Penrose} as to whether a black hole in equilibrium in a box is the same thing as a white hole in equilibrium in a box, or not, etc. What we can infer from our conjectures is that, if Hawking is right, and a black hole in a box is the same thing as a white hole in a box, then we should picture this as in Figure \ref{fig2} and not as in Figure \ref{fig1} (and not as Regions I and II of Figure \ref{fig1}).  (We remark that where we have referred to time-reversal symmetry above and in the previous paragraph, we should probably, more correctly, refer to a suitable generalization to quantum gravity of the notion of PCT symmetry.  Of course we admit that at present there is perhaps no such clear notion except possibly in a scattering-theoretic framework.)

What we have written above about the case of a Schwarzschild black hole in a box is based on our stress-energy singularity and almost-singularity conjectures and the latter has, as we have seen, a stronger statement when there is also an image box in the left wedge.   Thus it may still seem, perhaps, to leave open the possibility that an equilibrium quantum state for a spherical black hole in a spherical box could yet have an approximate classical spacetime description resembling Figure \ref{fig1} but with no box wall in the position of the dashed hyperbola.  However in the recent work on the one-mirror (/one box) case mentioned in Section \ref{Sect:instabilityargument} Umberto Lupo and I have obtained a result which tells us that there can be no equilibrium state on this latter spacetime, even in the context of quantum field theory in curved spacetime.  The details will appear elsewhere \cite{Kay-Lupo}.   This result thus strengthens our above conclusion that the only viable approximate description of a black hole in equilibrium in a spherical box in terms of a classical spacetime must resemble Figure \ref{fig2}.  

\subsection{Relation with the matter-gravity entanglement hypothesis}

The question that interests us most about all this is the following:  In the quantum case, when gravity is switched off,  Figure \ref{fig1} does make sense and we know that the HHI state of matter $<$\ref{Note:BWSewell}$>$, restricted to Region I, is thermal (in particular, impure)  because, in the overall (pure) HHI state, Region I is entangled with Region III.   When we now switch on gravity and also allow for matter and gravity perturbations,  then, assuming our conjectures and assumptions above (in the PCT-symmetric case) we will have to replace the Figure \ref{fig1} picture by the Figure \ref{fig2} picture.   It seems reasonable to then expect, though, that the quantum state of matter in Region I of Figure \ref{fig2} will (outside the `brick wall' region) still resemble the, thermal, restriction of the original HHI state to Region I of Figure \ref{fig1}.  But now, on these assumptions, there is no `left-wedge' Region III for this to be entangled with!  Yet, we do not wish to give up our usual assumption that the total state of a physical system is pure.  So one will essentially be forced to conclude that the matter fields in Region I of Figure \ref{fig2} must be entangled with something else.  What can that something else be?  \textit{The most natural answer would seem to be the gravitational field!}  This conclusion 
would seem to fit perfectly with the author's earlier \textit{matter-gravity entanglement hypothesis}  \cite{Kay1, Kay2, Kay-Abyaneh, Kay3, Kay4, Kay5} and in particular with the \textit{entanglement picture of black hole equilibrium} discussed in \cite{Kay3} (and, prior to that, in \cite{Kay1} and \cite{Kay-Abyaneh}).

In \cite{Kay-Ortiz} (to which we must refer for all details) we argue that an understanding of quantum Schwarzschild-AdS along these lines would seem to go some way towards resolving the puzzle \cite{ArnsdorfSmolin} raised by Arnsdorf and Smolin about how Maldacena holography \cite{Maldacena1} can be reconciled with Rehren's \textit{algebraic holography} \cite{Rehren}.

\subsection{The absence of a similar instability for plain AdS}
\label{Sect:AdSabsence}

We remark that, while we have argued for an instability of the maximally extended Schwarzschild-AdS spacetime, and the resulting invalidity of the Figure \ref{fig1} picture for that spacetime, we would not expect a similar instability for plain AdS even though that spacetime does contain a bifurcate Killing horizon (for an appropriate choice of Killing vector) and so four regions in some ways analogous to Regions I, II, III and IV of Figure 1. (See e.g.\ what we call `1+1 dimensional BTZ' in \cite{Kay-Ortiz} and e.g.\ \cite{Czechetal,  ParikhSamantray} for its higher dimensional plain-AdS counterpart.  Other relevant related papers include e.g.\ \cite{Casinietal, delaFuenteSundrum}.) The reason this latter situation is different is because, in the latter case, the enclosure (i.e.\ conformal boundary) is asymptotically timelike (rather than asymptotically null as depicted in Figure 1).   So we would not expect the sort of pile-up that we found above for right-going plane waves near the horizon ${\cal H}_B$, and we would not expect there to be finite initial energy classical waves which develop a singularity in their stress-energy tensor after reflecting off the conformal boundary.

\subsection{Connection with the work of Mathur and of Chowdhury et al}

Since a first version of this paper appeared, other authors \cite{AveryChowdhury, Mathur} (see also \cite{Chowdhury, ChowdhuryParikh}) have come to the conclusion, on seemingly quite different grounds,  that, in Schwarzschild-AdS, the quadruple wedge picture of Figure 1 should be replaced by a picture consisting of a right wedge Region I similar to our Figure 2 together with a left wedge Region III (the reflection in the vertical axis of Figure 2)  -- but with no future or past wedges (no Regions II and IV).   And they also share with the present paper the conclusion that there must be a non-classically describable quantum region near the horizons of Regions I and III as we have it in our Figure 2 (and in its reflection) -- understood by them in terms \cite{Mathur} of `fuzzballs'.   However, these papers appear to differ from us in considering overall pure quantum states on the union of Regions I and III which are entangled between these two regions, whereas, as we have argued three paragraphs above, we claim here that the state which corresponds to a physical asymptotically AdS black hole in equilibrium is a pure state of quantum gravity on the right wedge Region I alone which is entangled between matter and gravity.   If one wants to imagine a Region III being involved in the story, it would be, for us, a Region III which neither interacts with, nor is entangled with, Region I and therefore may as well not be there.  

\section{Notes}
\label{Sect:notes}

\begin{enumerate}

\item
\label{Note:moot} 
It is a moot point whether we should assume there to be an image box wall at the same $r=R$, on the other side of the Schwarzschild throat, as represented by the dashed hyperbola in Region III in Figure \ref{fig1}; we will mostly assume, below, that there is such an image box wall but we remark that our classical stress-energy tensor singularity for initial data with finite energy tails which we argue for here will obviously also occur when there is only one box wall and we expect the same to be true, in a suitable sense, in the quantum theory. 

\item
\label{Note:Deltar}  
One expects $\Delta r$ (cf.\ \cite{tHooft, Mukohyama-Israel}) to be of the order of the square of the Planck length divided by $8M$ since with that value, 
the metrical distance from the inner hyperbola to the horizon crossing point will be close to the Planck length.

\item
\label{Note:energy} To spell out what we mean by the ``usual notion of energy'' -- say of an initially right-moving solution $\phi$ supported in Regions IV, I and II:  Adopting the notation of Section \ref{Sect:mirrorsing}, such a solution will take the form $\phi(u,v)=f(u) + g(v)$ where $f$ is supported on $(-\infty, 0)$ and where $g(v)=-f(-1/v)$ so as to ensure that vanishing boundary conditions are satisfied on the mirror in Region I.  What we mean by this notion of energy of such a $\phi$ is then the energy of the solution, $\phi_M(u,v)=f(u)$, to the 1+1 dimensional wave equation on ordinary Minkowski space (i.e.\ without any mirrors) -- the Lorentz frame in which it is calculated being that defined by the condition that its $t$ and $x$ coordinates are related to $u$ and $v$ by $u=t-x$, $v=t+x$.   Explicitly, it is simply $\int_{-\infty}^0 (df(u)/du)^2\, du$.   We remark that it is the same thing as the integral over some
(one-dimensional) Cauchy surface, $\cal C$ for Region IV of
$T_{ob}N^b$ with respect to the induced volume element on the surface, where $N^a$ denotes the future pointing unit normal to the surface.  For example, we could choose, as such a Cauchy surface, a hyperbola $t^2-x^2=\mathrm{const}^2$, $t<0$ -- see Figure \ref{fig4}.  Parametrizing such a surface by $x \in (-\infty, \infty)$, this will obviously amount just to the integral over this surface of $({\partial\phi \over\partial x})^2 \, dx$.   

\item
\label{Note:chiandonemirror}  
The function, $f(u)=e^{-i\omega u}$ for $u < 0$ and $0$ for $u \ge 0$, in the paragraph after Equation (\ref{gfrelation}) fails to be smooth (i.e.\ $C^\infty$) -- being, of course, discontinuous at $u=0$. However we can remedy this by taking, instead, $f(u)=e^{-i\omega u}\chi(u)$, say, where $\chi$ is a smooth approximation to a step function, which takes the value 1 on some interval $(-\infty,-\epsilon]$ ($\epsilon>0$), zero on  $[0, \infty)$ and goes smoothly from 1 to 0 in the interval $[-\epsilon, 0]$, whereupon we will have $g(v)=-e^{i\omega /v}$ for $0<v < \epsilon^{-1}$ and still have $g(v)=0$ for $v\le 0$.

\smallskip

Our temporary assumption that the left-hand mirror is absent before Equation (\ref{fp}) is made partly so as to avoid having to multiply $f_p(u)$ by such a $\chi(u)$ to get a smooth $f(u)$ supported in $(-\infty, 0)$; clearly for $v\in [0, 1/\epsilon]$, the $g_p(v)$ of (\ref{onegfrelation}) is the same as we would have with both mirrors present were $f_p(u)$ to be multiplied by such a $\chi(u)$.  Indeed, $g_p(v)$ in Regions I and II depends only on the values of $f_p(u)$ when $u$ is negative, while its behaviour near ${\cal H}_B$, and in particular, the behaviour of $T_{vv}$ near ${\cal H}_B$ are entirely determined by the tail in $f_p(u)$ at large negative $u$ -- which goes like $u^{-p}e^{-i\omega u}$.

\smallskip

Finally we note that when both mirrors are present, we can define a solution ${\phi'}_p^B(u,v)$ to be $f_p(u) + g'_p(v)$ where $f_p(u)$ is as in (\ref{fp}) and $g'_p(v)=-f_p(-1/v)$:   We can think of this as having (characteristic) `initial' value $f_p(u)$ on the null-line ${\cal H}_B$.   The solution can then be thought of as emerging from both sides of ${\cal H}_B$, reflecting to the future off the right-hand mirror and to the past off the left-hand mirror and piling up both above and below ${\cal H}_B$.  This is relevant to Section \ref{Sect:quantumversion} where we discuss of the quantum theory with two mirrors.

\item
\label{Note:naturalQ} A natural question is how what we might call the `initial energy', i.e.\ $\int_{-\infty}^\infty (d\mathrm{Re}[f(u)]/du)^2 du$, of $\phi_p^B(u,v)$  -- or rather the half of this quantity that we can think of as the initial energy (see Endnote $<$\ref{Note:energy}$>$) of the restriction of $\phi_p^B(u,v)$ to Region IV -- of the solution $\phi_p^B(u,v)$ relates to the integral of its energy density, $T_{tt}$, over a constant-$t$ Cauchy surface, say for negative $t$, which (say in the case of two mirrors) begins at the left-hand mirror and ends at the right-hand mirror.  The answer is that, because of the contribution of the reflected part of the wave, $g_p(v)$, on the part of such a constant-$t$ surface between the horizon and the right-hand mirror, the two quantities are unequal.   Moreover this latter integral of $T_{tt}$ will not be conserved in time and, depending on the value of $p$, may not even be finite.  

\smallskip

What is of course true $<$\ref{Note:energy}$>$ is that, for any Cauchy surface for Region IV (for example the hyperbola defined in Endnote $<$\ref{Note:energy}$>$ and illustrated in Figure \ref{fig4}) with unit normal $N$, what we have called (half) the initial energy of $\phi_p^B(u,v)$ will equal the integral over that surface, with respect to the induced (i.e.\ from the Minkowski metric) (one-dimensional) volume element of that surface, of the stress energy tensor, $T_{ab}$, of $\phi_p^B(u,v)$ contracted with the vectors $\delta^a_0$ and $N$.     We note that this result may be regarded as satisfactorily justifying our notion of `initial energy' especially in view of the fact (pointed out also in Section \ref{Sect:silver}) that, in the presence of both mirrors, the Cauchy problem for the entire region between the two mirrors, is obviously well-posed for initial data on any such Cauchy surface for Region IV.  

\smallskip

Nevertheless, the fact that our initial energy is not the same thing as the integral of $T_{tt}$ over a constant $t$ line connecting the two mirrors may be regarded as an unsatisfactory aspect of our stress-energy singularity result in Section \ref{Sect:mirrorsing}.   It would be desirable to have a result along the lines:  \textit{An initially arbitrarily small perturbation on a constant $t$ line connecting the mirrors at early times leads to a singular (or maybe arbitrarily large) stress energy tensor near the horizon to the future.}   This -- in addition to the reasons given in Section \ref{Sect:silver} -- is one of our motivations for seeking, and obtaining, there, an alternative to that stress-energy singularity result in terms of initial data which have compact support on suitable Cauchy surfaces.   In particular, the ``more significantly different alternative to our `silver plated stress-energy almost-singularity result'' of our main text' explained in Endnote  $<$\ref{Note:straightalternative}$>$ can be seen to be along just such lines.

\item
\label{Note:initial}  
Note that, to qualify as a small perturbation, we would also want the solution to have, aside from finite total initial energy, no singularities in the energy density at any point in the interior of Region IV.   This is obviously satisfied by the solutions $\phi_p^B(u,v)$.

\item 
\label{Note:realmirrors}
While, as we pointed out in the main text, the total work done on the mirror can be finite and the force needed to keep it on its trajectory can be at all times finite, it might be objected that the singularity in $T_{vv}$, and also the `silver-plated stress-energy almost singularity result' argued for in Section \ref{Sect:silver}, are some steps removed from being results about physically realistic mirrors.  First of all, our result is in 1+1 dimensions.   Secondly, one might think it more physically realistic to ask about modified mirror trajectories which are initially and finally inertial and only accelerate for a finite amount of proper time.   Temporarily leaving aside the former issue it is easy to convince oneself that  for such modified mirror trajectories, one should be able, with suitable arbitrarily small smooth and even compactly supported Cauchy data, to still make $T_{vv}$ as large as we like at spacetime events (near where the horizon would be were the mirror motion not to be modified) by letting the interval of proper time for which the mirror accelerates uniformly be sufficiently long.  Moreover by having an image mirror in the left wedge with a similarly modified trajectory, it is reasonable to expect it to be possible, by an obvious extension of the arguments in Section \ref{Sect:silver}, to make the scalar quantity $T_{uu}T_{vv}$ as large as we like at some events near where the bifurcation point would be were the mirror trajectories not to be modified.   An inevitable issue which arises now is that, the longer we make these proper-time intervals, the more work we expect  will have to be done at the start and/or end of the accelerated portion of the mirror trajectory (/trajectories) because of vacuum friction \cite{Davies, WangUnruh}.   The point is that we expect that one or other of the portions of each of the mirror trajectories which interpolate between inertial and uniformly accelerated motion, and during which the mirror-motion is non-uniformly accelerated, will be responsible for (ever-larger) amounts of particle creation.   The created particles will be radiated off to infinity and we expect increasing amounts of (irretrievable) work would need to be done on the mirror to supply these radiated particles with their energy.   (Additional work will of course have to be done to accelerate the mirror(s) because of its (/their) mass. However, this can be made as small as we like by making the mirror mass(es) small enough.)   If we now turn to 1+3 dimensions, it seems reasonable to assume that similar results will continue to hold for say sufficiently large, say flat, mirrors transversal to their direction of motion, and for electromagnetic, instead of scalar perturbations.

Thus, it seems reasonable to conclude that, in principle, one could bring about spacetime regions with arbitrarily large $T_{uu}T_{vv}$, and hence, in view of Einstein's equations, with arbitrarily high curvature invariant, $R_{ab}R^{ab}$, by a suitable device involving mirrors which accelerate for a finite interval of their proper time and a suitable small perturbation in the electromagnetic field (i.e.\ suitable pulse of light), provided one expends sufficient energy to overcome the vacuum friction mentioned above.  

\item
\label{Note:KWtechnicalities} 
Note also that in \cite{Kay-Wald} when definining $S_A$ and $S_B$ in generic spacetimes with bifurcate Killing horizons, as explained in the \textit{Note Added in Proof} there, one inevitably gets involved with spaces of solutions which are only differentiable a certain finite number of times.

\item
\label{Note:checksuitable} 
To check that ${\phi'}_p^B(u)$ (defined in Endnote $<$\ref{Note:chiandonemirror}$>$) is suitable for $p > 1/2$, we may use the fact (cf.\ \cite{Kay-Wald}) that the right hand side of (\ref{KBKB}) is equal to $2\int_0^\infty k\tilde f_1^*(k)\tilde f_2(k) dk$ where $\tilde f(k)=\int_{-\infty}^\infty f(u)e^{-iku}du$.  One easily sees that, for $p > 1/2$, both ${\phi'}_p^B$ and its $u$-derivative belong to $L^2(\mathbb{R})$.  Hence, both $\tilde{\phi'}_p^B$ and $k\tilde{\phi'}_p^B$ also belong to $L^2(\mathbb{R})$ (the $\mathbb{R}$ now being `momentum space').  The right hand side of $\langle K{\phi'}_p^B|K{\phi'}_p^B\rangle$ is  then twice the inner product $\langle k\tilde{\phi'}_p^B|\tilde{\phi'}_p^B\rangle$ in this latter $L^2$ space which, by Cauchy-Schwartz is less than or equal to $\langle \tilde{\phi'}_p^B|\tilde{\phi'}_p^B\rangle^{1/2}\langle k\tilde{\phi'}_p^B|k\tilde{\phi'}_p^B\rangle^{1/2}$, which is finite.

\smallskip

Actually, one can see by direct computation that ${\phi'}_p^B(u)$ is also suitable for
$p=1/2$.  

\smallskip

(I thank Umberto Lupo for pointing out that a previous version of this footnote only worked for $p>1$ and for a helpful discussion on how to fix it.)

\item
\label{Note:easycoherent}  
In fact for any linear scalar field theory there will be a notion of symplectically smeared fields (see Section 3 of \cite{Kay-Wald}) and, by the standard commutation relations for these, we will have $e^{i\hat\phi(\phi)}\hat\phi(x)e^{-i\hat\phi(\phi)}=\hat\phi(x) + \phi(x)$.  Given a `vacuum' state determined by a one-particle structure, $(K,H)$, we will then easily have that the expectation value of the renormalized stress-energy tensor in our coherent state is equal to the sum of the stress-energy tensor of the classical solution $\phi$ and the vacuum expectation value of the renormalized stress-energy tensor.  In our case, as we mention in the main text, the latter is zero.  Note that, in the paragraph after Equation (\ref{mirrordiverge}), we apply this general result to what we call there ``the total renormalized/classical `initial' total energy on the B horizon'' of $\phi'_B(u,v)$ and what we mean by this is $\int T_{uu} du$ where $T_{uu}$ is the renormalized/classical stress-energy tensor of $\phi'_B(u,v)$ and the integral is over the line $v=0$ (i.e.\ over ${\cal H}_B$).

\item
\label{Note:characteristicversion}  
A technically different but physically closely related conjecture (which we also make) is that, say, for a scalar field on our 1+1-Minkowski mirror system times a 2-torus, the solution in the right wedge Region I determined by the characteristic data $\phi_p(u)$ for $u < 0$ ($u$ being affine parameter on the $u <0$ half of the B-horizon) say times a constant or times $e^{im\theta+in\phi}$, where $\theta$ and $\phi$ are the torus angles, will have a singular $T_{vv}$ on the same half-horizon and similarly for Schwarzschild in a box and Schwarzschild-AdS with $e^{im\theta+in\phi}$ replaced by a general spherical harmonic.

\item
\label{Note:Rindlerfootnote}   
Note that since Rindler time-evolution is a symmetry of our 1+1-Minkowski system with (one or both) mirrors, Rindler energy will be conserved.

\item
\label{Note:history} 
That an observer passing from Region IV to Region I will see an infinite amount of history can be easily verified whenever the enclosure is asymptotically null and, in particular, holds for our accelerated mirror in Minkowski space and also for Schwarzschild in a box and for Schwarzschild-AdS.  However one can see that it won't hold for plain AdS where (see the penultimate paragraph of Section \ref{Sect:Discussion}) although there are counterpart regions to Region IV and Region I, the conformal boundary is not asymptotically null.

\item
\label{Note:any}  
In fact, for the Cauchy horizon of  Reissner-Nordstr\"om, the stress-energy singularity holds for \textit{any} smooth compactly supported initial Cauchy data on a suitable initial surface $<$\ref{Note:any}$>$.   This is in contrast to our `silver-plated 
almost-singularity result' of Section \ref{Sect:silverconjecture}.

\item
\label{Note:boost} 
To help the reader verify the various statements made in the main text around the reference to this end-note,  note first that when specifying Cauchy data for right-going solutions on some initial spacelike surface, we only need to specify the first piece of Cauchy data -- equal to the restriction of the solution to the surface -- since the second piece of data (equal, say, to the future-pointing normal derivative of the solution restricted to the surface) is then determined by the condition that the solution be right-going (and similarly for left-going solutions).     If, for the purposes of discussing Cauchy data for right-going solutions, we coordinatize our initial surface, $t^2-x^2=\mathrm{const}^2$, $t<0$, by $u$ then, if we denote the first piece of Cauchy data by the function $f$ of $u$, then we mean, by a negative boost of this data, the data whose first piece is $f^{\mathrm{boosted}}$ of $u$ where $f^{\mathrm{boosted}}(u)=f(e^{-\tau}u)$ say.    We then easily have, e.g., that $T_{uu}$ of our boosted solution at $u$ is equal to $e^{-2\tau}$ times $T_{uu}$ of our original solution at $e^{-\tau} u$ while $T_{vv}$ of our boosted solution, after reflection at the right-hand mirror, at $v$ will be $e^{2\tau}$ times $T_{vv}$ of our original solution, after reflection at the right-hand mirror, at $e^\tau v$ etc.

\item
\label{Note:infinitescalar} 
Note that a similar argument to that we gave in the main text for an arbitrarily large (scalar) $T_{uu}T_{vv}$ in our `silver-plated stress-energy almost singularity result' will easily show that our original singularity result  with finite-energy tails may be strengthened to include the statement that, in the case of two mirrors, such initial data can be chosen (by taking it to consist of a sum of suitable left-going and right-going initial data in Region IV) so that the scalar quantity  $T_{uu}T_{vv}$ is \textit{singular} at the bifurcation point.

\item
\label{Note:straightalternative} 
We note that, as a mildly different alternative to our `silver-plated stress-energy almost singularity result'  we might (cf.\ $<$\ref{Note:characteristicversion}$>$) have formulated it in terms of characteristic Cauchy data on the union of the negative-$u$ part of ${\cal H}_B$ and the negative-$v$ part of ${\cal H}_A$ rather than in terms of Cauchy data on our initial surface $t^2-x^2=\mathrm{const}^2$, $t<0$ .  We also note the following more significantly different alternative to our `silver-plated stress-energy almost singularity result' of our main text which is based on Cauchy data on constant $t$ lines rather than on the $t<0$ branch of the single spacelike hyperbola 
$t^2-x^2=\mathrm{const}^2$ of our main text:   Consider a (countably) infinite family of equally spaced (in $t$) constant negative-$t$ lines, marching towards the past, in (the fixed Lorentz frame of) Figure \ref{fig1} which each join our two mirrors and have non-empty intersections with Regions III, IV and I and consider a sequence of classical solutions, the first of which has arbitrarily small smooth compactly supported data on the intersection of the future-most of these lines with Region IV consisting of the sum of a right-going solution located just to the left of ${\cal H}_B$ and a left-going solution located just to the right of ${\cal H}_A$ and the $n$th term of which consists of the sum of right-going and left-going solutions whose Cauchy data are translations in $t$ and $x$ of each set of Cauchy data on the first line so that each is similarly located -- i.e.\ the right-going data is located just to the left of ${\cal H}_B$ and the left-going data is located just to the right of ${\cal H}_A$ -- on the $n$th line.   Then it is easy to see that, by taking $n$ sufficiently large, the $n$th solution will have a $T_{vv}$ as large as we like near ${\cal H}_B$ and a $T_{uu}$ as large as we like near ${\cal H}_A$ and a $T_{uu}T_{vv}$ as large as we like near the bifurcation point.   This alternative `silver-plated stress-energy almost singularity result' has, however, the disadvantage that it will not readily generalize, from our 1+1-Minkowski situation with two mirrors, to a conjecture about Schwarzschild in a box and Schwarzschild-AdS since (a) (Kruskal etc.) $t$-translations are not symmetries (b) in these latter spacetimes, we can only draw such constant $t$ lines to the future of the past singularity.

\item
\label{Note:unstable}
Of course it is well-known there is another sense in which a black-hole in equilibrium in a box is unstable.  Namely \cite{HawkingBHT} it is thermodynamically unstable due to a negative specific heat.  However this is a separate issue from the new issues we raise in the present paper and not one on which we wish to comment here.   What we would emphasize though is that, as pointed out by Hawking and Page \cite{HawkingPage}, for a suitable range of values of the mass and cosmological constant, this thermodynamics instabibility is absent for Schwarzschild-AdS black holes and, as we have argued, our new sort of instability obviously still holds also for those cases.

\item
\label{Note:BWSewell} 
In the present paper, we have modelled matter by our massless linear scalar field in 1+1 dimensional Minkowski space but it is known that the thermal/entanglement results mentioned in this paragraph generalize to an arbitrary quantum field theory in Minkowski space of any dimension \cite{Sewell} and we expect them to generalize beyond that to a wide class of spacetimes with bifurcate Killing horizons \cite{Kay-Wald} for which HHI-like states exist too.

\end{enumerate}

\ack
 
I wish to thank an anonymous referee of the paper \cite{Kay-Ortiz} for asking a question which stimulated some of the work reported here.   I thank Eli Hawkins, Atsushi Higuchi, Hugo Ferreira and Jorma Louko for helpful remarks and criticisms of an earlier version of this paper.   I thank Umberto Lupo for a critical reading of that earlier version and also for assistance with, and checks of, many of my calculations and also for assistance with Endnote $<$\ref{Note:checksuitable}$>$ .  I also wish to thank Chris Fewster for a valuable discussion and, in particular, for a specific suggestion (indicated in a parenthetical remark above) which helped me to make the present `silver-plated stress-energy almost-singularity result' considerably stronger than a previous version.  I also thank Borun Chowdhury for drawing to my attention the references \cite{AveryChowdhury, Mathur} and for an interesting discussion on the connection between the work in those references and the present work.


\section*{References}
\begin{thebibliography}{99}



\bibitem{Maldacena1} Maldacena, J.M.: The large N limit of superconformal field theories and supergravity. Adv. Theor. Math. Phys. {\bf 2}, 231 (1998) [also published as, Int.\ J.\ Theor.\ Phys. {\bf 38}, 1113 (1999)].  arXiv:hep-th/9711200

\bibitem{Maldacena2} Maldacena, J.M.: Eternal black holes in anti-de Sitter. JHEP {\bf 4}, 021 (2003). arXiv:hep-th/0106112

\bibitem{tHooft} 't Hooft, G.: On the quantum structure of a black hole.  Nucl.\ Phys.\ B {\bf 256} 727 (1985)

\bibitem{Mukohyama-Israel} Mukohyama, S., Israel, W.: Black holes, brick walls and the Boulware state. Phys.\ Rev.\ D {\bf 58} 104005
(1998).  arXiv:gr-qc/9806012

\bibitem{Press-Teuk} Press, W., Teukolsky, S.: Floating orbits, superradiant scattering and the black-hole bomb. Nature {\bf 238} 211 (1972)

\bibitem{Eardley} Eardley, D.M.: Death of white holes in the early universe. Phys.\ Rev.\ Lett.\ {\bf 33} 442 (1974)

\bibitem{BlauGuth} Blau, S.K., Guth, A.H.: The stability of the white hole horizon. (1989) (manuscript submitted to the Gravity Research Foundation.  Available at
http://gravityresearchfoundation.org/pdf/awarded/1989/blau\_guth.pdf)

\bibitem{Blau} Blau, S.K.: Dray 't Hooft geometries and the death of white holes. Phys.\ Rev.\ D {\bf 39} 2901 (1989)

\bibitem{Lake} Lake, K.: Reissner-Nordström-de Sitter metric, the third law, and cosmic censorship. Phys.\ Rev.\ D {\bf 19} 421 (1979)

\bibitem{WaldRamaswamy} Wald, R.M., Ramaswamy, S.: Particle production by white holes. Phys.\ Rev.\ D {\bf 21} 2736 (1980)

\bibitem{HawkingBHT} Hawking, S.W.: Black holes and thermodynamics. Phys.\ Rev.\ D {\bf 13} 191 (1976)

\bibitem{HawkingPage} Hawking, S.W., Page, D.: Thermodynamics of black holes in anti 
de-Sitter space.  Commun.\ Math.\ Phys. {\bf 87} 577 (1973) 

\bibitem{Davies} Davies, P.C.W.: Quantum vacuum friction.  J. Opt. B: Quantum Semiclass. Opt. {\bf 7} S40–S46 (2005) 

\bibitem{WangUnruh} Wang, Q., Unruh, W.G. Motion of a mirror under infinitely fluctuating quantum vacuum stress.  Phys.\ Rev.\ D {\bf 89}, 085009 (2014) arXiv:1312.4591

\bibitem{Hartle-Hawking} Hartle, J.B., Hawking, S.W.:  Path-integral derivation of black-hole radiance.  Phys.\ Rev.\ D {\bf 13} 2188 (1976)

\bibitem{Israel} Israel, W.: Thermofield dynamics of black holes.  Phys.\ Lett.\ A {\bf 57} 107 (1976)

\bibitem{Wightman} Wightman, A.S.: Introduction to some aspects of the relativistic dynamics of quantum fields.  In: 1964 Carg\`ese Lectures in Theoretical Physics: High Energy Electromagnetic Interactions and Field Theory. ed. M. L\'evy.  Gordon and Breach, New York 1967

\bibitem{Fulling-Ruijsenaars} Fulling, S.A., Ruijsenaars, S.N.M.: Temperature, periodicity and horizons. Phys.\ Rep. {\bf 152} 135 (1987)

\bibitem{KayNantes} Kay, B.S.: Application of linear hyperbolic PDE to linear quantum fields in curved spacetimes: especially black holes, time machines and a new semi-local vacuum concept. Journées Equations aux Dérivées Partielles IX-1 (2000) arXiv:gr-qc/0103056

\bibitem{Kay-Wald} Kay, B.S., Wald, R.M.: Theorems on the uniqueness and thermal properties of stationary, nonsingular, quasifree states on spacetimes with a bifurcate Killing horizon. Phys.\ Rep. {\bf 207} 49-136 (1991).  (Note that the uniqueness result in this paper was later strengthened in Kay B.S.: Sufficient conditions for quasifree states and an improved uniqueness theorem for quantum fields on space–times with horizons.  J.\ Math.\ Phys.\ {\bf 34}, 4519 (1993).)

\bibitem{KayPhD}  Kay, B.S.: Quantum Fields in Time-Dependent Backgrounds and in Curved Space-times.  University of London PhD thesis (1977)

\bibitem{Jaffe-Ritter} Jaffe, A., Ritter, G.: Reflection postivity and monotonicity.  J.\ Math.\ Phys.\ {\bf 49} 052301 (2008). 
arXiv:0705.0712

\bibitem{Birrell-Davies} Birrell, N.D., Davies, P.C.W.: Quantum Fields in Curved Space. Cambridge University Press, Cambridge (1982)

\bibitem{KayCasimir} Kay, B.S. The Casimir effect in quantum field theory.  (\textit{Original title} The Casimir effect without magic.)  Phys.\ Rev.\ D {\bf 20} 3052 (1979)

\bibitem{Rindler} Rindler, W.  Kruskal space and the uniformly accelerated frame.  Am.\ J.\ Phys. {\bf 34} 1174 (1966)

\bibitem{Simpson-Penrose} Simpson, M., Penrose, R.: Internal instability in a Reissner-Nordström black hole. Int.\ J.\ Theor.\ Phys. {\bf 7} 183 (1973) 

\bibitem{Hawking-Ellis} Hawking, S.W., Ellis, G.F.R.: The Large Scale Structure of Space-Time. Cambridge University Press, Cambridge (1973)

\bibitem{Chandra-Hartle} Chandrasekhar, S., Hartle, J.B.: On crossing the Cauchy horizon of a Reissner-Nordstrom black-hole Proc.\ R.\ Soc.\ A {\bf 384} 301 (1982)

\bibitem{Hiscock} Hiscock, W.A.: Stress-energy tensor near a charged, rotating, evaporating black hole. Phys.\ Rev.\ D {\bf 15} 3054 (1977)

\bibitem{Dafermos1} Dafermos, M.: Stability and instability of the Cauchy horizon for the spherically symmetric Einstein-Maxwell-scalar field equations.  Annals of Mathematics {\bf 158} 875 (2003)

\bibitem{Dafermos2} Dafermos, M.:  Stability and Instability of the Reissner-Nordstrom Cauchy horizon and the problem of uniqueness in general relativity. Contemp.\ Math.\ {\bf 350} 99 (2004). arXiv:gr-qc/0209052

\bibitem{Kay-Lupo} Kay, B.S., Lupo, U.: \textit{Work in progress.}

\bibitem{Sewell} Sewell, G.L.: Quantum fields on manifolds: PCT and gravitationally induced thermal states. Ann.\ Phys.\ (NY) {\bf 141} 201 (1982)

\bibitem{Hawking-Penrose} Hawking, S.W., Penrose, R.: The Nature of Space and Time. Princeton University Press, Princeton (1996, 2010)

\bibitem{Kay1} Kay, B.S.: Entropy defined, entropy increase and decoherence understood, and some black-hole puzzles solved (1998). arXiv:hep-th/9802172

\bibitem{Kay2} Kay, B.S.: Decoherence of macroscopic closed systems within Newtonian quantum gravity. Class.\ Quant.\ Grav.\ {\bf 15} L89-L98 (1998). arXiv:hep-th/9810077

\bibitem{Kay-Abyaneh} Kay, B.S., Abyaneh, V.: Expectation values, experimental predictions, events and entropy in quantum gravitationally decohered quantum mechanics (2007).
arXiv:0710.0992

\bibitem{Kay3} Kay, B.S. On the origin of thermality (2012). arXiv:1209.5125

\bibitem{Kay4} Kay, B.S.: Modern foundations for thermodynamics and the stringy limit of black hole equilibria (2012). arXiv:1209.5085

\bibitem{Kay5} Kay, B.S.: More about the stringy limit of black hole equilibria (2012). arXiv:1209.5110

\bibitem{Kay-Ortiz} Kay, B.S., Ort\'iz, L.: Brick walls and AdS/CFT. J.\ Gen.\ Rel.\ Grav. {\bf 46} 1727 (2014).  arXiv:1111.6429

\bibitem{ArnsdorfSmolin} Arnsdorf, M., Smolin, L.: The Maldacena conjecture and Rehren duality (2001). arXiv:hep-th/0106073

\bibitem{Rehren} Rehren, K.-H.: Algebraic holography. Ann. Henri Poin car\'e {\bf 1} 607 (2000). arXiv:hep-th/9905179

\bibitem{Czechetal} Czech, B., Karczmarek, J.L., Nogueira, F., Van~Raamsdonk, M.: Rindler quantum gravity. Class.\ Quantum\ Grav. {\bf 29} 235025 (2012). arXiv:1206.1323

\bibitem{ParikhSamantray} Parikh, M., Samantray, P.: Rindler-AdS/CFT (2012).
arXiv:1211.7370

\bibitem{Casinietal} Casini H., Huerta M., Myers R.C.: Towards a derivation of holographic entanglement entropy. JHEP {\bf 05} 036 (2011).
arXiv:1102.0440

\bibitem{delaFuenteSundrum} de~la~Fuente A., Sundrum R.: Holography of the BTZ black hole, inside and out (2013). arXiv:1307.7738

\bibitem{AveryChowdhury} Avery S.G., Chowdhury B.D.: No holography for eternal AdS black holes (2013). arXiv:1312.3346

\bibitem{Mathur} Mathur S.: What is the dual of two entangled CFTs? (2014). arXiv:1402.6378

\bibitem{Chowdhury} Chowdhury B.D.: Limitations of holography (2014). arXiv:1405.4292

\bibitem{ChowdhuryParikh} Chowdhury B.D., Parikh M.K.: When UV and IR Collide: Inequivalent CFTs From Different Foliations Of AdS (2014). arXiv:1407.4467

\end{thebibliography}
\end{document}